\documentclass[prd,reprint,longbibliography,nofootinbib,superscriptaddress]{revtex4-2}

\usepackage[english]{babel}
\usepackage[utf8x]{inputenc}
\usepackage[T1]{fontenc}

\usepackage{tikz}
\usepackage{amsmath}
\usepackage{amssymb}
\usepackage{mathtools}
\usepackage{xcolor}
\usepackage[colorlinks,allcolors=violet]{hyperref}
\usepackage{cleveref}
\usepackage{graphicx}
\usepackage{booktabs}
\usepackage[acronym]{glossaries}
\usepackage{smartdiagram}
\usesmartdiagramlibrary{additions}
\usepackage{float}
\usepackage{mathrsfs}
\usepackage{bm}
\usepackage{dsfont}

\usepackage{physics}
\usepackage{siunitx}
\usepackage{soul}

\newcommand{\Gfilter}{{\mathbf{G}}}
\newcommand{\filter}{{\mathbf{F}}}
\newcommand{\sample}{{\mathbf{S}}}
\newcommand{\tdi}{{\mathbf{C}}}

\newcommand{\Delay}[1]{{\hat{\mathbf{D}}_{#1}}}

\newcommand{\fDelay}[1]{{\hat{\mathcal{D}}_{#1}}}

\newcommand{\dotfDelay}[1]{{\dot{\hat{\mathcal{D}}}_{#1}}}

\newcommand{\delay}[1]{{\mathbf{D}_{#1}}}

\newcommand{\fdelay}[1]{{\mathcal{D}_{#1}}}

\newcommand{\dotdelay}[1]{{{\mathbf{\dot D}}_{#1}}}

\DeclareSIUnit\year{yr}

\DeclareMathOperator{\sinc}{sinc}
\DeclareMathOperator{\rect}{rect}

\newacronym{gw}{GW}{gravitational wave}
\newacronym{esa}{ESA}{European Space Agency}
\newacronym{lisa}{LISA}{the Laser Interferometer Space Antenna}
\newacronym[longplural=movable optical sub-assemblies]{mosa}{MOSA}{movable optical sub-assembly}
\newacronym{inrep}{INREP}{initial noise-reduction pipeline}
\newacronym{tdi}{TDI}{Time-delay interferometry}
\newacronym{tt}{TT}{travel time}
\newacronym{prn}{PRN}{pseudo-random noise}
\newacronym{ppr}{PPR}{proper pseudo-range}
\newacronym[longplural=power spectral densities]{psd}{PSD}{power spectral density}
\newacronym[longplural=amplitude spectral densities]{asd}{ASD}{amplitude spectral density}
\newacronym{fir}{FIR}{finite impulse response}
\newacronym{uso}{USO}{ultra-stable oscillator}

\begin{document}

\title{Laser noise residuals in LISA from onboard processing and time-delay interferometry}
\author{Martin Staab}
\affiliation{Max Planck Institute for Gravitational Physics (Albert Einstein Institute), D-30167 Hanover, Germany}
\affiliation{Leibniz Universit\"at Hannover, D-30167 Hanover, Germany}
\author{Marc Lilley}
\affiliation{SYRTE, Observatoire de Paris, Universit\'e PSL, CNRS,
Sorbonne Universit\'e, LNE, 61 avenue de l’Observatoire 75014 Paris, France}
\author{Jean-Baptiste Bayle}
\affiliation{University of Glasgow, Glasgow G12 8QQ, United Kingdom}
\author{Olaf Hartwig}
\affiliation{SYRTE, Observatoire de Paris, Universit\'e PSL, CNRS,
Sorbonne Universit\'e, LNE, 61 avenue de l’Observatoire 75014 Paris, France}
\affiliation{Max Planck Institute for Gravitational Physics (Albert Einstein Institute), D-30167 Hanover, Germany}
\affiliation{Leibniz Universit\"at Hannover, D-30167 Hanover, Germany}

\date{\today}




\begin{abstract}
\Gls{tdi} is a crucial step in the on-ground data processing pipeline of the \gls{lisa}, as it reduces otherwise overwhelming laser noise and allows for the detection of \glspl{gw}. This being said, several laser noise couplings have been identified that limit the performance of \gls{tdi}. First, on-board processing, which is used to decimate the sampling rate from tens of \si{\mega\Hz} down to a few \si{\Hz}, requires careful design of the anti-aliasing filters to mitigate folding of laser noise power into the observation band. Furthermore, the flatness of those filters is important to limit the effect of the flexing-filtering coupling. Secondly, the post-processing delays applied in \gls{tdi} are subject to ranging and interpolation errors. All of these effects are partially described in the literature. In this paper, we present them in a unified framework and give a more complete description of aliased laser noise and the coupling of interpolation errors. Furthermore, for the first time, we discuss the impact of laser locking on laser noise residuals in the final \gls{tdi} output. To verify the validity of the analytic \gls{psd} models we derive, we run numerical simulations using \texttt{LISA Instrument} and calculate second-generation \gls{tdi} variables with \texttt{PyTDI}. We consider a setup with six independent lasers and with locked lasers (locking configuration N1-12). We find that laser locking indeed affects the laser noise residual in the \gls{tdi} combination as it introduces correlations among the six lasers inducing slight modulations of the \glspl{psd} compared to the case of six independent lasers. This implies further studies on laser noise residuals should consider the various locking configurations to produce accurate results.

\end{abstract}
\maketitle
\glsresetall

\section{Introduction}%
\label{sec:introduction}

\Gls{lisa} is a space mission led by the \gls{esa}, expected to be launched in the 2030s. Its goal is to detect \glspl{gw} in a frequency band ranging from \SI{E-4}{\hertz} to \SI{1}{\hertz}~\cite{amaroseoane2017laser}. High precision interferometric measurements will be made via the exchange of laser beams among three spacecraft orbiting the Sun and separated by 2.5 million kilometers, in order to determine the variations in the distance between free-falling test masses aboard each spacecraft to picometer precision. In these measurements, the laser phase noise is the primary noise source and is 8 orders of magnitude larger than the \gls{gw} signals that one hopes to detect.  \gls{tdi} is a data processing technique that combines the \gls{lisa} measurements to construct virtual equal-arm two-beam interferometers in order to reduce the laser phase noise to levels sufficiently low such that \glspl{gw} become detectable~\cite{Tinto:1999yr,Armstrong_1999}.
In \gls{tdi} the measurements are time-delayed by multiples of the \gls{lisa} arm lengths and combined in a specific scheme to achieve laser phase noise reduction.  Second-generation \gls{tdi}, which is the current baseline laser phase noise reduction strategy for \gls{lisa}, applies to the case in which the arm lengths of the \gls{lisa} constellation evolve slowly and linearly in time~\cite{Shaddock:2003dj,Tinto:2003vj}. In second generation \gls{tdi},
laser phase noise is strongly suppressed and the residual ia fundamentally limited by the arm length mismatch of the virtual interferometer \cite{Bayle:2021mue}.

There exist other approaches to perform laser phase noise suppression.  In \gls{tdi}-$\infty$~\cite{Vallisneri:2020otf}, the observables that cancel laser phase noise are obtained numerically by solving for the null space of the design matrix, i.e., the way the various noise sources enter the interferometric measurements, for an arbitrary time dependence of the arm lengths. The likelihood function that is used in \gls{gw} source parameter estimation can then be written directly in the time domain in terms of the \gls{lisa} interferometric measurements without having to reformulate the entire problem in terms of algebraically defined \gls{tdi} variables.  While the study in~\cite{Vallisneri:2020otf} was limited to an idealized toy model with a single Michelson interferometer, the authors of~\cite{Houba:2023wmm} applied \gls{tdi}-$\infty$ to the full \gls{lisa} constellation with time-evolving arms. Computationally, \gls{tdi}-$\infty$ has the drawback that it requires the storage and manipulation of very large matrices.

In~\cite{Baghi:2021lmk}, starting from the interferometric measurements and for non-evolving \gls{lisa} arms, the authors first form a matrix of integer-delayed measurements, which they decompose using principal component analysis (PCA) into high and low variance components. The latter correspond to the components for which the laser phase noise is significantly suppressed.  This approach, dubbed ``automated Principal Component Interferometry'', or aPCI, is formulated in both the time and frequency domain. In~\cite{Baghi:2022wdd}, the same authors extend this approach to the case of time-evolving arms. Note that in~\cite{Baghi:2022wdd}, aPCI is shown not to perform as well as second-generation \gls{tdi} in suppressing the laser phase noise.

While approaches such as \gls{tdi}-$\infty$ and aPCI offer some interesting perspectives for a flexible data-driven formulation of \gls{tdi}, ``traditional'' \gls{tdi} can be formulated analytically. It is therefore tractable, better understood, and exact analytic transfer functions exist to describe the instrumental noise residuals present in the \gls{tdi} variables.  For instance, secondary noises such as for example test-mass acceleration noise and optical metrology system noise are dealt with in~\cite{Nam:2022rqg}, clock noise is studied in~\cite{Hartwig:2020tdu,Hartwig:2022wcs}, and tilt-to-length coupling in~\cite{Paczkowski:2022nrt}. Note also that laser noise coupling residuals were discussed previously in \cite{Bayle:2018hnm,Hartwig:2021dlc}. It is also worth stressing that a good understanding of the noise content in the final \gls{tdi} output is crucial to characterize the performance of \gls{lisa} and guide the design of the instrument and that data analysis and parameter estimation will require accurate noise models  in order to work reliably, making these studies particularly relevant for \gls{lisa}.

In addition to the analytic and numerical studies available in the literature, there exist several hardware demonstrators that test various aspects of \gls{tdi} experimentally. The LISA interferometry test-bed \citep{PhysRevLett.104.211103}, while it could not reproduce the signal delays in a realistic way, did demonstrate for the first time that using first generation \gls{tdi}, both the laser and clock noise could be suppressed by 9 and 4--5 orders of magnitude, respectively.
In UFLIS \citep{Mitryk_2010,PhysRevD.86.122006}, using electronic phase delay units allowing for time-varying delays of the laser phases, the authors were able to demonstrate the efficacy of second-generation \gls{tdi}. In more ecent work \citep{PhysRevD.105.042009}, the Hexagon experiment demonstrated that clock synchronization can be achieved to sufficient accuracy to match the \gls{lisa} requirements. Moreover, the authors find residual laser noise after \gls{tdi}-like processing due to flexing-filtering, aliasing, and interpolation error. The LISA on table (LOT) experiment \citep{Gruning:2013gca} (for recent progress see \citep{leonvidal:2023}) is an electro-optical setup aiming primarily at testing the laser noise suppression performance of \gls{tdi}. In~\citep{leonvidal:2023}, the validity of second-generation \gls{tdi} was demonstrated for linearly evolving \gls{lisa} arms. It was also shown using an analytic model that the residual noise could be explained by the cascade integrator comb filtering and the decimation stages that are applied to the data.
  
\begin{figure*}
    \centering
    \includegraphics[width=14cm]{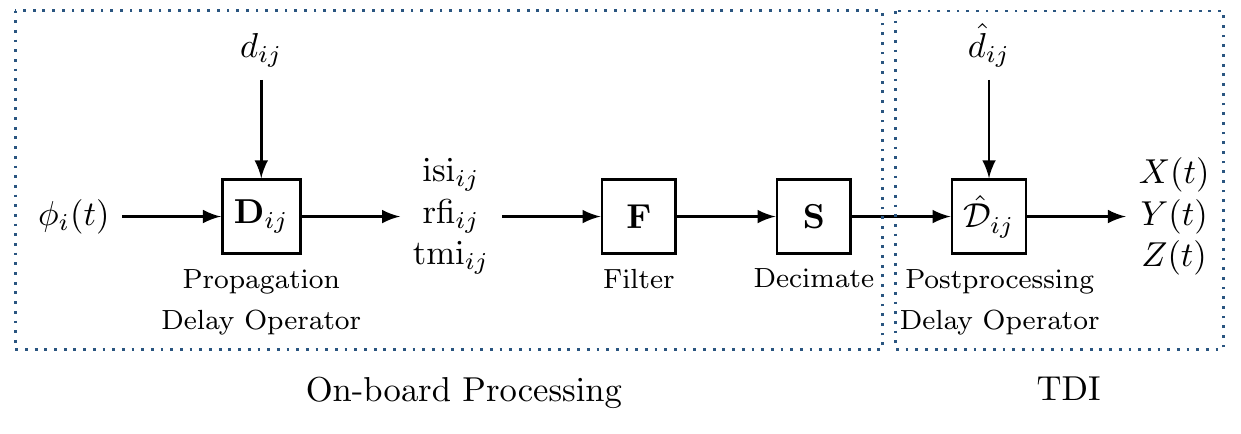}
    \caption{Illustration of the formation of interferometric beatnote phases and successive on-board processing steps on the left and post-processing \gls{tdi} on the right. The propagation and post-processing delay are defined in \cref{sec:preamble} and \cref{sec:processing_residuals}, respectively} 
    \label{fig:flowchart}
\end{figure*}

In this paper, we study the coupling of laser noise residuals in standard \gls{tdi}.  We focus on the residual laser noise due to systematic effects and neglect most other noise sources. The one exception is noise in the ranging measurements which are used as delays in TDI, and which in principle couple to laser frequency noise. Following \cite{Hartwig:2022wcs}, we assume this noise source will be strongly suppressed to the level of the highly-precise sideband interferometer readouts, such that its impact on the laser noise reduction is minor. We still include it in so as to have a more complete description of the post-processing delay.  We consider the effect of on-board processing (i.e. filtering and decimation), and the influence of \gls{tdi} which uses post-processing delay operations that are subject to ranging and interpolation errors, see \Cref{fig:flowchart}. We compute analytic formulae for all laser noise residuals induced by these processing steps and compare those to numerical simulations obtained using \texttt{LISA Instrument} and \texttt{PyTDI}.  We do so for six independent lasers and also for locked lasers in the N1-12 locking scheme~\cite{heinzel2018,Bayle:2022okx}.

The laser noise residuals induced by the processing steps we derive below are already partially described in the literature. The flexing-filtering effect arising from the non-commutativity of the filtering and delay operation was previously discussed in~\cite{Bayle:2018hnm}. The impact of ranging errors, i.e. modulation noise and ranging biases, in \gls{tdi} was derived in~\cite{Hartwig:2022wcs}, while a preliminary models for the aliased laser noise due to decimation and interpolation errors were derived in~\cite{Hartwig:2021dlc}.  Here, we gather these results in a unified framework and check them against the most up-to-date instrumental setups. In addition, we include the effect of laser locking in our models, an important feature of \gls{lisa} that was previously neglected. We show that laser locking can amplify the laser noise residual in the \gls{tdi} combinations, so that future performance studies on laser noise coupling should consider the influence of the various locking configurations that are available. Finally, we correct the preliminary models presented in~\cite{Hartwig:2021dlc}, which did not show perfect agreement with simulation results.

The paper is organised as follows.  In \cref{sec:preamble}, we introduce the interferometric measurements available in \gls{lisa}, the notion of discretely sampled time series and how filtering and decimation apply to discretized data.  In \cref{sec:commutator_residuals} we discuss the residual laser noise induced by the non-commutativity of the on-board filtering and decimation operations with the propagation delays. The interpolation and ranging residuals, which both arise from the on-ground processing steps, are discussed in \cref{sec:processing_residuals}.  In \cref{sec:michelson}, we apply the results of the previous sections to the case of the second-generation \gls{tdi} Michelson variables for six independent lasers and for locked lasers.  Additional details are given in four appendices.

\section{Interferometric measurements}%
\label{sec:preamble}

\Gls{lisa} produces two main interferometric measurements per \gls{mosa} relevant for laser noise reduction\footnote{The split interferometry configuration involves a third interferometer, the test-mass interferometer, which is not relevant for the purpose of this study}. Those are the inter-spacecraft and reference interferometers given by
\begin{subequations}
    \begin{align}
        \mathrm{isi}_{ij}(t) &= \sample\filter\big( \delay{ij} \phi_{ji}(t) - \phi_{ij}(t) \big), \label{eq:isi}\\
        \mathrm{rfi}_{ij}(t) &= \sample\filter\big( \phi_{ik}(t) - \phi_{ij}(t) \label{eq:rfi}\big).
    \end{align}
\label{eq:phasemeter}
\end{subequations}
Each measurement represents the beatnote phase formed by two laser beams, whose phases are denoted by $\phi$ and labeled by the index pair $ij$. Here, we follow the conventions in \cite{Bayle:2021mue}, where $i$ denotes the hosting spacecraft and $j$ the spacecraft from which \gls{mosa} $ij$ receives light. The inter-spacecraft interferometer $\mathrm{isi}_{ij}$ tracks the phase difference between the distant laser $\phi_{ji}$ that is propagated to the local \gls{mosa} and the local laser $\phi_{ij}$. Beam propagation is equivalent to applying the delay operator $\delay{ij}$ defined as
\begin{align}
    \delay{ij}\phi_{ji}(t) = \phi_{ji}(t - d_{ij}(t)).
\end{align}
Here, $d_{ij}(t)$ is the pseudo-range, which includes the light travel time in some chosen global frame and any reference frame transformation accounting for the fact that the phases $\phi_{ji}$ are defined in their respective reference frame, $j$. In order to relate phases on the left and right \glspl{mosa} on each spacecraft\footnote{Each spacecraft is equipped with two \glspl{mosa}. The left-handed \gls{mosa} on spacecraft~$i$ refers to the one facing spacecraft~$i+1$, while the right-handed \gls{mosa} refers to the one facing spacecraft~$i-1$ (indices ranging from 1 to 3 cyclic).}, the reference interferometer $\mathrm{rfi}_{ij}$ combines the local and the adjacent lasers.

It is useful to decompose the total laser phase $\phi_{ij}(t)$ or the frequency $\nu_{ij}(t) = \dot\phi_{ij}(t)$ into two variables because, as we shall see below, the instrumental and data pre-processing effects we will describe (see the diagram of \cref{fig:flowchart}) couple differently to the phase ramp $\phi_{ij}^o(t)$ and any in-band fluctuations $\phi_{ij}^\epsilon(t)$. We thus write
\begin{align}
    \phi_{ij}(t) &= \phi_{ij}^o(t) + \phi_{ij}^\epsilon(t) \label{eq:phi}, \\
    \nu_{ij}(t) &= \nu_{ij}^o(t) + \nu_{ij}^{\epsilon}(t),
    \label{eq:nu}
\end{align}
where $\nu_{ij}^o(t) = \dot\phi_{ij}^o(t)$ describes any slowly-varying drifts around the central laser frequency $\nu_0 = \SI{281.6}{\tera\Hz}$ and $\nu_{ij}^\epsilon(t) = \dot\phi_{ij}^\epsilon(t)$ accounts for any rapidly varying random fluctuations. This in-band part is dominated by laser frequency noise with an \gls{asd} $\sqrt{S_{\dot p}}=\SI{30}{\Hz\per\sqrt{\Hz}}$.

The beatnote phases of the interferometers are read out using a digital phase locked loop running at $\SI{80}{\mega\Hz}$. Multiple decimation stages reduce the sampling rate down to \SI{4}{\Hz} in order to produce the final data streams telemetered to ground. Each decimation stage consists of an anti-aliasing filter, $\filter$, and a downsampling stage, $\sample$, which downsamples the data by an integer factor. In this work, we compare the analytic models that we derive with the most recent \gls{lisa} simulation codes, which run at rates that are much lower than the \SI{80}{\mega\Hz} quoted above, and thus only use a single decimation stage. This being said, the results obtained in this paper can easily be generalized to multiple decimation stages.

We shall express signals in continuous time so as to be compatible with the recent literature on \gls{tdi}. However, the application of \gls{fir} filters, decimation, and interpolation requires some notion of discretely sampled time series. We will therefore make use of the Whittaker-Shannon interpolation formula~\cite[e.g.,][]{jenkins_spectral_1968},
\begin{equation}
    x(t) = \sum_{n=-\infty}^\infty \sinc(f_s t - n) \cdot x_n,
    \label{eq:whittaker_shannon}
\end{equation}
which reconstructs the continuous time signal $x(t)$ from discrete samples $x_n$.\\


Let us first describe the onboard processing, which consists of the application of a \gls{fir} filter and a decimation stage.  A \gls{fir} filter is equivalent to a discrete convolution of the input time series $x_n$ with filter taps $h_m$,
\begin{equation}
y_n = \sum_m h_m \cdot x_{n-m}.
    \label{eq:filtering_discrete}
\end{equation}
We use \cref{eq:whittaker_shannon} to represent the output $y_n$ in continuous time and find
\begin{equation}
y(t) = \int_\mathbb{R}\! \underbrace{\sum_m h_m \delta(\tau - m T_s)}_{h_\filter(\tau)} x(t - \tau)\,\mathrm{d} \tau = \filter x(t), \label{eq:filtering_continuous}
\end{equation}
where we introduce the integral over the Dirac-delta distribution $\delta(t)$ to shift the time argument of $x(t)$.
It follows that the application of a \gls{fir} filter is equivalent to a continuous time convolution with the filter kernel $h_\filter(\tau)$ defined above.  Using the usual definition of the one-sided \gls{psd}, the \gls{psd} of the filtered process in \cref{eq:filtering_continuous} is given by
\begin{equation}
    S_y(f) = \left|\tilde h_\filter(f)\right|^2 S_x(f) = \tilde\filter S_x(f)
    \label{eq:filtering_psd}
\end{equation}
with $\tilde h(f)$ the Fourier transform of $h(t)$, while $S_x(f)$ is the \gls{psd} of $x(t)$.

Let us now discuss the decimation operator which we use to reduce the sampling rate by an integer factor $M$. On a discrete time grid, the resulting signal is given by
\begin{equation}
y_n = x_{n\cdot M}\,.
\label{eq:decimation_discrete}
\end{equation}
Again, we make use of \cref{eq:whittaker_shannon} to find the corresponding continuous time representation,
\begin{equation}
y(t) = \sample\, x(t) = \sum_{n=-\infty}^\infty \sinc(f_s t - n) \cdot x_{n\cdot M} .
\label{eq:decimation_continuous}
\end{equation}
Here, $f_s$ denotes the sampling rate {\it after} decimation, and $\sample$ symbolizes the action of the decimation operation in continuous time. The right-hand side of \cref{eq:decimation_continuous} is exactly equal to $x(t)$ if and only if it has a band limit that is less than the Nyquist rate after decimation, $f_n = f_s/2$.
Otherwise aliasing occurs, which folds power from frequencies above $f_n$ into the band $[0, f_n]$. 
This effect becomes apparent when looking at the corresponding one-sided \gls{psd},
\begin{equation}
    S_{y}(f) = \tilde\sample\, S_x(f) = \rect\left(\frac{f}{f_s}\right)\sum_{n=0}^{M-1} S^{(n)}_x(f)\,.
    \label{eq:sampling_psd}
\end{equation}
Here, $\tilde\sample\, S_x(f)$ is a shorthand notation representing the action of the decimation operator in Fourier space on the \gls{psd} of $x(t)$ and the rectangular function  is defined to be equal to zero for $|f| > f_n$ and equal to one for $|f| < f_n$ such that the decimated signal is band-limited up to the new Nyquist rate.  Finally, the $n^\mathrm{th}$ alias, $S^{(n)}_x(f)$, on the right-hand-side of \cref{eq:sampling_psd}, is given by
\begin{equation}
    S_x^{(n)}(f) =
    \begin{cases*}
        S_x(n f_n + f) & if $n$ is even,\\
        S_x((n + 1) f_n - f) & if $n$ is odd.
    \end{cases*}
    \label{eq:aliases}
\end{equation}
\Cref{eq:sampling_psd,eq:aliases} highlight the typical folding into band of any spectral component that resides at frequencies higher than the new Nyquist rate (up to the highest frequency $M f_n$, which corresponds to the Nyquist rate before decimation).

\section{Commutator Residuals}%
\label{sec:commutator_residuals}
When computing \gls{tdi} laser noise residuals, expressions of the form
\begin{equation}
    [A, B]\phi_{ij}(t) = A B \phi_{ij}(t) - B A \phi_{ij}(t) \label{eq:comm_general}
\end{equation}
appear, with operators $A$ and $B$ that act on the time series $\phi_{ij}(t)$. We call two operators non-commutative if \cref{eq:comm_general} is non-vanishing. As an example, the fundamental limit for laser noise suppression in \gls{tdi} can be described by a commutator of time-dependent delay operators, see e.g. \cite{Bayle:2021mue}. This is discussed in more detail in \cref{sec:michelson}).

Let us now outline how the filtering and sampling operators enter as additional commutators in \gls{tdi}.

The basic building block of every \gls{tdi} combination is the set of intermediary variables $\eta_{ij}$ (defined later in \cref{sec:michelson}). In the idealized case for which none of the data processing steps depicted in \cref{fig:flowchart} are considered (i.e., $\filter=\sample=1$ and $\mathcal{\hat{D}}_{ij}=\mathbf{D}_{ij}$) the variable $\eta_{ij}$ simplifies to the difference between a local laser phase with a distant laser phase delayed by the light travel time between spacecraft $i$ and $j$.  For the left-handed \glspl{mosa}, one has
\begin{equation}
    \eta_{ij}(t) = \delay{ij} \phi_{jk}(t) - \phi_{ij}(t).\label{eq:eta_generic}
\end{equation}
The expression for the right-handed \glspl{mosa} is similar but the indexing is different. Under these conditions, i.e., with the specific algebraic form of \cref{eq:eta_generic}, the fundamental laser noise limit is the the usual delay commutator (see e.g. \cref{eq:X_delay_commutator}).

If we instead insert the interferometric measurements of \cref{eq:phasemeter} into the definition of $\eta_{ij}$ given in \cref{eq:eta}, the resulting expression cannot be recast into the algebraic form of \cref{eq:eta_generic}. This is because of the order in which the filtering, decimation, and delay operations arise in \cref{eq:isi}. If one introduces the following commutator into \cref{eq:phasemeter}\,,
\begin{equation}
     [\sample\filter,\delay{ij}]\phi_{ji}(t) = (\sample\filter)\delay{ij}\phi_{ji}(t) - \delay{ij}(\sample\filter)\phi_{ji}(t),
\end{equation}
the delay operator $\delay{ij}$ switches places with the decimation stage $\sample\filter$ such that \cref{eq:isi} becomes
\begin{equation}
    \mathrm{isi}_{ij}(t) = \delay{ij} \sample\filter \phi_{ji}(t) - \sample\filter \phi_{ij}(t) + [\sample\filter, \delay{ij}]\phi_{ji}(t).
\label{eq:phasemeter_comm}
\end{equation}
This expression has the same algebraic form as \cref{eq:eta_generic} if we make the replacement $\phi_{ij} \rightarrow \sample\filter\phi_{ij}$, with the exception that the inter-spacecraft interferometric measurement now also contains the commutator $[\sample\filter, \delay{ij}]\phi_{ji}$. This commutator comes as an additional residual in the final \gls{tdi} expressions (this is reminiscent of the way readout noise enters the measurements, see~\cite{Nam:2022rqg}).

Using common commutator rules we can further split $[\sample\filter, \delay{ij}]\phi_{ji}$ into a filter-delay commutator and a decimation-delay commutator
\begin{equation}
    [\sample\filter,\delay{}]\phi(t) =  \sample[\filter,\delay{}]\phi(t) + [\sample,\delay{}]\filter\phi(t), \label{eq:comm_sample_filter_delay}
\end{equation}
and compute explicit analytic expressions for each contribution, see \cref{ssec:flexing_filtering_coupling} and \cref{ssec:fractional_delay_sampling_coupling}.

\subsection{Flexing-filtering coupling}%
\label{ssec:flexing_filtering_coupling}
Let us derive the contribution coming from the commutator $[\filter,\delay{}]$, first described in~\cite{Bayle:2018hnm} and dubbed ``flexing-filtering coupling''. We assume that the delay $d(t)$ is slowly varying over the filter length and that its first derivative $\dot d(t)$ is small. We use \cref{eq:filtering_continuous}, and expand $[\filter, \delay{}]\phi(t)$ at leading order in $\dot d(t)$ to find
\begin{equation}
    [\filter,\delay{}] \phi(t) \simeq \dot d(t)\delay{}\Gfilter\nu(t).
\label{eq:FFcommutator}
\end{equation}
where $\Gfilter$ is a filter operator defined similarly to $\filter$ in \cref{eq:filtering_continuous} with $h_\Gfilter(\tau) = \tau \cdot h_\filter(\tau)$. 

Because of the orbital dynamics of the \gls{lisa} constellation, $\dot d(t)$ will vary on time scales of several months and, as a result, so will the level of the laser noise residual introduced by the flexing-filtering coupling. However, over sufficiently short observation times we can assume $\dot d$ to be constant. In such a case the \gls{psd} of \cref{eq:FFcommutator} reads
\begin{equation}
S_{\delta\phi}^{[\filter,\delay{}]}(f) = \dot d^2 \left|\frac{1}{2\pi}\frac{\mathrm{d}\tilde h_\filter(f)}{\mathrm{d}f}\right|^2 S_\nu(f). \label{eq:psd_comm_filter_delay}
\end{equation}
For longer observation times, one can use the maximum value of $\dot d$ as given by current predictions for the orbital dynamics of \gls{lisa} in order to derive an upper bound for this \gls{psd}.
\subsection{Decimation-delay commutator}%
\label{ssec:fractional_delay_sampling_coupling}
The second commutator appearing in \cref{eq:comm_sample_filter_delay} is the commutator of the decimation and delay operations,
\begin{equation}
    [\sample,\delay{}] \phi(t) = \sample\delay{}\phi(t) - \delay{}\sample\phi(t).
    \label{eq:comm_sample_delay}
\end{equation}
Those operations do not commute due to the non-linear nature of the decimation process.
The \gls{psd} of this expression can be derived using the definition in \cref{eq:decimation_continuous}. The different aliases that are folded in band are modulated by a sine-squared factor. We obtain 
\begin{equation}
    S_{\delta\phi}^{[\sample,\delay{}]}(f) = 4 \cdot \sum_{n=1}^{M-1} c_n(d) \cdot S_\phi^{(n)}(f) ,\label{eq:psd_comm_sample_delay}
\end{equation}
where the modulating factor $c_n$ is given by
\begin{equation}
    c_n(d) =
        \begin{cases*}
            \sin[2](\pi f_s d \frac{n}{2}) & for $n$ even,\\
            \sin[2](\pi f_s d \frac{n+1}{2}) & for $n$ odd.
        \end{cases*}
\end{equation}
In the special case where the delay $d$ becomes an integer multiple of the sampling time $T_s=1/f_s$ decimation and delay operations commute and the residual becomes zero. For a time-varying delay, this particular residual is non-stationary because its power is modulated as $d$ evolves in time. To account for this, we later consider only the upper bound obtained for $c_n=1$ for all $n$,
\begin{equation}
    S_{\delta\phi}^{[\sample,\delay{}]}(f) \le 4 \cdot \sum_{n=1}^\infty S_\phi^{(n)}(f).
\label{eq:psd_comm_sample_delay_sim}
\end{equation}
This bound is independent of the delay, and corresponds to the case of full anti-correlation between $\sample\delay{}\phi(t)$ and $\delay{}\sample\phi(t)$ in \cref{eq:comm_sample_delay}.

\subsection{Comparison with numerical simulations}%
\begin{figure}
    \centering
    \includegraphics{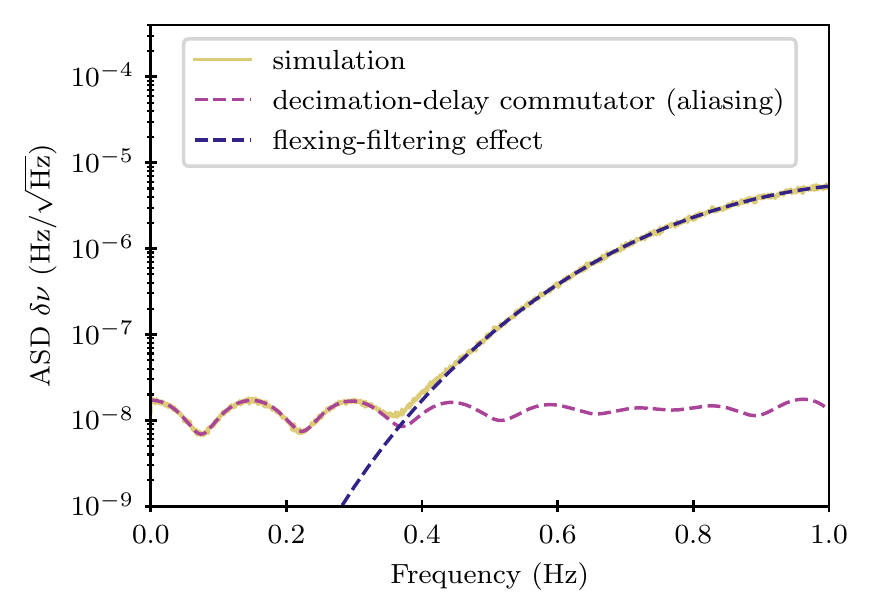}
    \caption{Commutator residuals from filtering and decimation. We compare the numerically simulated data (yellow) against analytic models (dashed lines). In purple we show the model for the coupling of the decimation-delay commutator (aliasing) and in indigo the flexing-filtering effect.}
    \label{fig:toymodel_commutator}
\end{figure}

The numerical simulations that are used in this work to validate the analytic models are performed in units of frequency in order to preserve numerical precision\footnote{A beatnote frequency of several \si{\mega\Hz} results in a rapidly increasing phase ramp. This would require a large number of significant digits to deal with its dynamic range ($\sim\SI{e14}{cycles}$ after \SI{1}{\year} at a required precision of at least \si{\micro cycles} which results in 20 significant digits). On the other hand, its derivative, the beatnote frequency, stays roughly constant and a double precision float (approximately 16 significant digits) is sufficient to represent it. Here, the \si{\micro cycles} requirement with a knee at \SI{2}{\milli\Hz} corresponds to a precision of approximately \SI{20}{\nano\Hz} which gives 15 orders of magnitude when considering a \SI{20}{\mega\Hz} beatnote frequency.}. As shown in~\cite{Bayle:2021mue}, any delay operation on frequency data can be represented by the usual shift of the argument and a multiplicative Doppler factor,
\begin{equation}
    \dotdelay{} \nu(t) = (1 - \dot d(t)) \cdot \nu(t - d(t)).
    \label{eq:dotdelay}
\end{equation}
We can then easily rewrite the commutator given in \cref{eq:comm_sample_filter_delay} in terms of frequency data by replacing every occurrence of the delay operator $\delay{}$ by its Doppler equivalent. It reads
\begin{equation}
    [\sample\filter,\dotdelay{}] \nu(t) = \sample\filter\dotdelay{} \nu(t) - \dotdelay{}\sample\filter \nu(t).
    \label{eq:comm_sample_filter_delay_freq}
\end{equation}
Here, we need to account for the Doppler factor to cancel laser noise to first order. However, we find that it only has a negligible impact on the laser noise residual, and we can write for the \glspl{psd}
\begin{subequations}
    \begin{align}
        S_{\delta\nu}^{\sample[\filter,\dotdelay{}]}(f) &\simeq \tilde\sample \left((2\pi f)^2 \dot d^2 \left|\frac{1}{2\pi}\frac{\mathrm{d}\tilde h_\filter(f)}{\mathrm{d}f}\right|^2 S_\nu(f)\right), \\
        S_{\delta\nu}^{[\sample,\dotdelay{}]\filter}(f) &\simeq 4 \sum_{n=1}^\infty c_n(d) \left(\tilde\filter S_\nu\right)^{(n)}\!(f) .
    \end{align}
    \label{eq:psd_comm_sample_filter_delay_freq}
\end{subequations}
We note that for the full commutator in \cref{eq:comm_sample_filter_delay}, we need to apply $\sample$ to the flexing-filtering contribution and account for the fact that laser noise is filtered prior to passing it through the decimation-delay commutator. In our numerical implementation, the effect of decimating the flexing-filtering residual was negligible compared to the in-band contribution. Nevertheless, we include it for completeness as it strongly depends on the filter design.

In order to test the validity of the theoretical model given by \cref{eq:psd_comm_sample_filter_delay_freq}, we compare it to numerical simulations that generate time series corresponding to \cref{eq:comm_sample_filter_delay_freq}. We consider white laser frequency noise in $\nu(t)$ with an \gls{asd} of \SI{30}{\Hz\per\sqrt{\Hz}}\footnote{In reality, we expect laser noise to increase towards lower frequencies. In this manuscript, we assume that it is a white noise to easily study the shape of various residuals. However, the residual transfer functions derived here hold for any laser noise spectrum.} and neglect the central laser frequency of \SI{2.816e14}{\Hz} (its coupling to the commutator is vanishing). The simulation is performed at a sampling rate of \SI{16}{\Hz} and is then decimated down to \SI{4}{\Hz}. Anti-aliasing is performed using an \gls{fir} designed according to \cref{app:filter_design}. The delay operator is modeled by numerical interpolation of the data using Lagrange polynomials (c.f. \cref{sec:processing_residuals}).

In \cref{fig:toymodel_commutator} we compare the simulated data against the analytical model presented in \cref{eq:psd_comm_sample_filter_delay_freq}. We use the Welch method from the \texttt{SciPy} package~\cite{Virtanen:2019joe} to estimate the \gls{psd} of the time series. The numerical result is explained by aliasing at low frequencies and by flexing-filtering at high frequencies.

\section{Post-processing residuals in TDI}%
\label{sec:processing_residuals}

The working principle of \gls{tdi} is to time-shift the recorded beatnote phase measurements and linearly combine them in order to reduce laser frequency noise in the resulting combinations. To achieve this we require an interpolation method and estimates of the aforementioned time shifts. In the following we denote a delay operation that is performed in on-ground data processing as $\fDelay{}$. This operator acts on a discrete time series and is therefore only a numerical approximation. Furthermore, time-shifting is performed with imperfect knowledge of the delay. The discreteness of the data and the error in ranging produce a residual with respect to the true propagation delay $\delay{}$. To distinguish between these effects, we write the residual due to the imperfections of $\fDelay{}$ as
\begin{equation}
    (\fDelay{} - \delay{})\phi(t) = (\fDelay{} - \Delay{}) \phi(t) + (\Delay{} - \delay{}) \phi(t) ,
    \label{eq:offline_delay_residual}
\end{equation}
where the first term on the right-hand side represents the interpolation residual and the second term the residual stemming from ranging errors. In the following we study both effects in detail.

\subsection{Interpolation residual}%

We follow~\cite{Shaddock:2004ua} and implement a post-processing delay in \gls{tdi} as an interpolation with a fractional delay filter.  Lagrange interpolation has demonstrated its suitability in this context, and we thus use it as the baseline in this work. In general, we can model post-processing delays as \gls{fir} filters. For  convenience, we split the delay into an integer shift $j$ and a fractional shift $\epsilon$ ranging from 0 to 1. The latter defines the coefficients of the interpolation kernel $k_n(\epsilon)$ with $-N/2 \leq n \leq N/2-1$, which is convolved with the discrete data samples $\phi_n$,
\begin{equation}
    (\fdelay{}\, \phi)_n = \sum_{m=-N/2}^{N/2-1} k_m(\epsilon) \cdot \phi_{n-j-m}.
    \label{eq:fdelay}
\end{equation}
Before the convolution of \cref{eq:fdelay} is performed, the data samples are shifted by the integer shift $j$. The above formula holds for interpolation kernels of even length $N$. For odd $N$, a similar expression can be derived. Using \cref{eq:filtering_continuous}, we find
\begin{equation}
    h_\fdelay{}(\tau) = \sum_{m=-N/2}^{N/2-1} k_m(\epsilon) \cdot \delta(\tau - (j+m) T_s) .
\end{equation}
We define the additional phase residual caused by the interpolation error as
\begin{equation}
    \delta \phi^{\fdelay{}}(t) = (\fdelay{} - \delay{}) \phi(t) = \delay{}\underbrace{(\delay{}^{-1}\fdelay{} - 1)}_{\bm{\Delta}} \phi(t)
\end{equation}
and use \cref{eq:filtering_psd} to derive the residual in terms of \gls{psd}. We find
\begin{equation}
    S_{\delta \phi}^\fdelay{}(f) = \underbrace{\left|\tilde h_\fdelay{}(f)e^{2\pi i f d} - 1\right|^2}_{\tilde{\bm{\Delta}}} S_\phi(f).
    \label{eq:interpolation_error_psd}
\end{equation}
In general, the interpolation kernel $k_m(\epsilon)$ has to be adjusted for every sample $n$ to account for time-dependent time shifts. Therefore, the flexing arms of \gls{lisa} will produce a non-stationary interpolation residual as the fractional delay $\epsilon$ scans through different values. At $\epsilon = 0$ and $\epsilon = 1$, the delay is a pure integer shift and the residual vanishes. Assuming that the worst case is obtained for $\epsilon=0.5$\footnote{A rigorous proof is needed to validate this assumption which was only found to be true {\it empirically}, in our work, and specifically for Lagrange interpolation.}, we can derive an upper bound for the residual induced by interpolation.

As suggested in~\cite{Shaddock:2004ua}, and already mentioned at the beginning of this section, a suitable interpolation method is Lagrange interpolation, and we use in this work.  The interpolation kernel $k_m$ is derived from fitting a Lagrange polynomial through a set of neighboring samples. This method is known for producing a maximally flat frequency response at DC and is therefore well suited for \gls{lisa} data processing, as it performs well over the entire \gls{lisa} band (\SIrange{e-4}{1}{\Hz}) when using high interpolation orders\footnote{Assuming a sampling rate of \SI{4}{\Hz}, typical filter lengths used for post-processing delays in \gls{tdi} are in the range 32 to 66.}. Alternative interpolation kernels are under study that use less coefficients and optimize their performance over the entire band (and not only at DC).

\subsection{Ranging residual}%
\label{sub:ranging_residual}
Any estimate of the delay $\hat{d}(t)$ differs from the true delay $d(t)$ by a ranging error $r(t)$. We define the corresponding delay operator as
\begin{equation}
    \Delay{}\phi(t) = \phi(t - d(t) - r(t)) \simeq \delay{}\phi(t) - r(t) \delay{}\dot \phi(t),
    \label{eq:hdelay_expansion}
\end{equation}
where we have assumed $r(t)$ to be small and performed a series expansion to first order. The ranging residual is then given by
\begin{equation}
    \delta \phi^{\Delay{}}(t) = (\Delay{} - \delay{}) \phi(t) = - r(t) \delay{}\nu(t) \label{eq:ranging_residual}
\end{equation}
with $\nu(t)=\dot\phi(t)$.
Similarly to laser phase or frequency, c.f. \cref{eq:nu}, we decompose the ranging error $r(t)$ into an out-of-band component $r^o(t)$ and an in-band component $r^\epsilon(t)$. Here, $r^o(t)$ absorbs ranging biases that are of the order of \SI{3}{\nano\s}~\cite{Hartwig:2022wcs} and might be slowly drifting. The in-band component $r^\epsilon(t)$ has a root-mean-squared value of $\sim\SI{100}{\femto\s}$ (assuming the \gls{psd} in~\cref{eq:modulation_noise}). Therefore, $r^\epsilon(t) \ll r^o(t)$ and we find as the prominent in-band contributions the coupling of laser noise to the ranging bias and the coupling of ranging noise to the \si{\mega\Hz} beatnote frequency
\begin{equation}
    r(t)\delay{}\nu(t) \simeq r^o(t)\delay{}\nu^\epsilon(t) + r^\epsilon(t)\delay{}\nu^o(t).
\end{equation}
For completeness, in \cref{app:ranging_laser_noise_coupling} we present the coupling of the stochastic in-band ranging error to laser frequency noise.

To simplify the calculations we assume the out-of-band components of the ranging error and the beatnote frequency to be constant. We can readily write down the \gls{psd} of \cref{eq:ranging_residual} as 
\begin{equation}
    S_{\delta \phi}^\Delay{}(f) = (r^o)^2 S_\nu(f) + (\nu^o)^2 S_r(f) .
    \label{eq:ranging_residual_psd}
\end{equation}

\subsection{Comparison with numerical simulations}%

\begin{figure}
    \centering
    \includegraphics{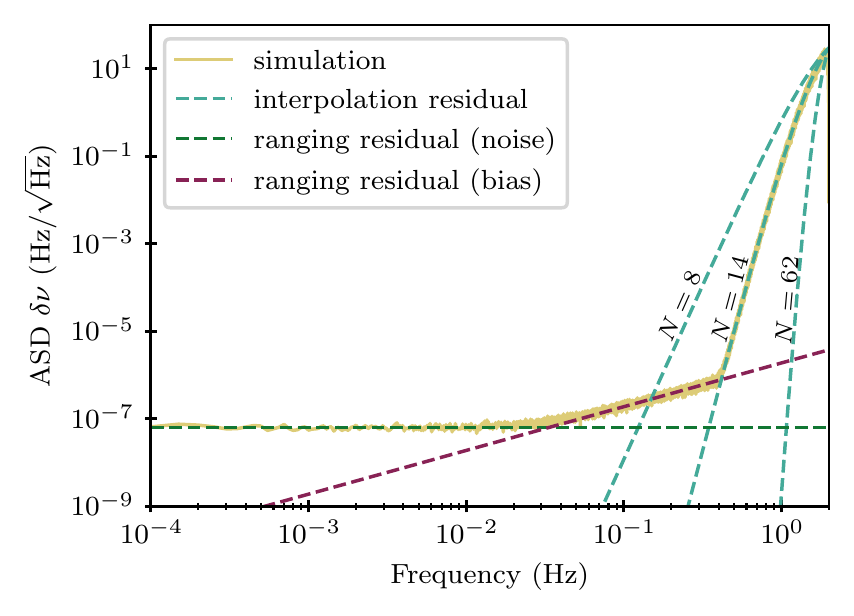}
    \caption{Processing residuals in application of the fractional delay filter containing ranging errors. We compare the numerically simulated data (yellow) against the analytical models for the ranging residual caused by ranging noise (green) and a constant bias (wine). Additionally, we plot models for the interpolation residual (teal) for different interpolation kernel lengths $N=8,14,62$.}
    \label{fig:toymodel_processing}
\end{figure}

Once again, we check the validity of our analytic model using simulations performed in units of frequency. 
Using \cref{eq:dotdelay}, we write down the expression corresponding to the left-hand-side of \cref{eq:offline_delay_residual} for frequency data as
\begin{equation}
    \delta \nu^{\fDelay{}}(t) = (1 - \dot{\hat d}(t)) \cdot \fDelay{}\nu(t) - (1 - \dot d(t)) \cdot \delay{}\nu(t) .\label{eq:delay_residual_freq}
\end{equation}
In order to simulate time series data corresponding to \cref{eq:delay_residual_freq}, we first generate a generic beatnote frequency with a constant offset of \SI{10}{\mega\Hz} and a white laser frequency noise component with an \gls{asd} of \SI{30}{\Hz\per\sqrt{\Hz}} at a sampling frequency of \SI{4}{\Hz}.  Both the post-processing delay $\fDelay{}$ and the propagation delay $\delay{}$ are implemented as fractional delay filters. To simulate the latter, we use a very high interpolation order ($N=502$), such that the interpolation error becomes negligible in comparison to that of the post-processing delay. The ranging error present in the post-processing delay is modeled by a bias $B = \SI{e-8}{\s}$, and ranging noise\footnote{We choose a red tilt for ranging noise to be easily distinguishable from coupling of the ranging bias to laser frequency noise. The level is comparable with realistic models of modulation noise discussed in \cref{sec:michelson}.} with an \gls{asd} of $\SI{e-15}{\s\per\sqrt{\Hz}} \left(\frac{\si{\Hz}}{f}\right)$ . The nominal value of the delay is taken to be equal to $d=\SI{8.125}{\s}$. This yields the worst case interpolation error, with a fractional part $\epsilon = 0.5$.

In \cref{fig:toymodel_processing} we compare the \gls{psd} of the numerical time series corresponding to \cref{eq:delay_residual_freq} with the analytic expressions for each of the two components of \cref{eq:interpolation_error_psd,eq:ranging_residual_psd} re-expressed for frequency data, i.e.,
\begin{subequations}
    \begin{align}
        S_{\delta \nu}^\fdelay{}(f) &= \tilde{\bm{\Delta}} S_{\nu}(f), \\
        S_{\delta \nu}^\Delay{}(f) &= (2\pi f)^2 \cdot S_{\delta \phi}^\Delay{}(f) .
    \end{align}\label{eq:delay_residual_freq_psd}
\end{subequations}
The interpolation residual (dashed teal) is strongly dependent on the length $N$ of the interpolation kernel. For this reason, we show the interpolation residual obtained for $N=8,\, 14$, and $62$. The post-processing delay used in the numerical implementation is performed with an interpolation order $N=14$. As shown in the figure, the model of \cref{eq:delay_residual_freq_psd} agrees with the data for all frequencies.

\section{Second-generation Michelson combinations}%
\label{sec:michelson}

\begin{figure}
    \centering
    \includegraphics{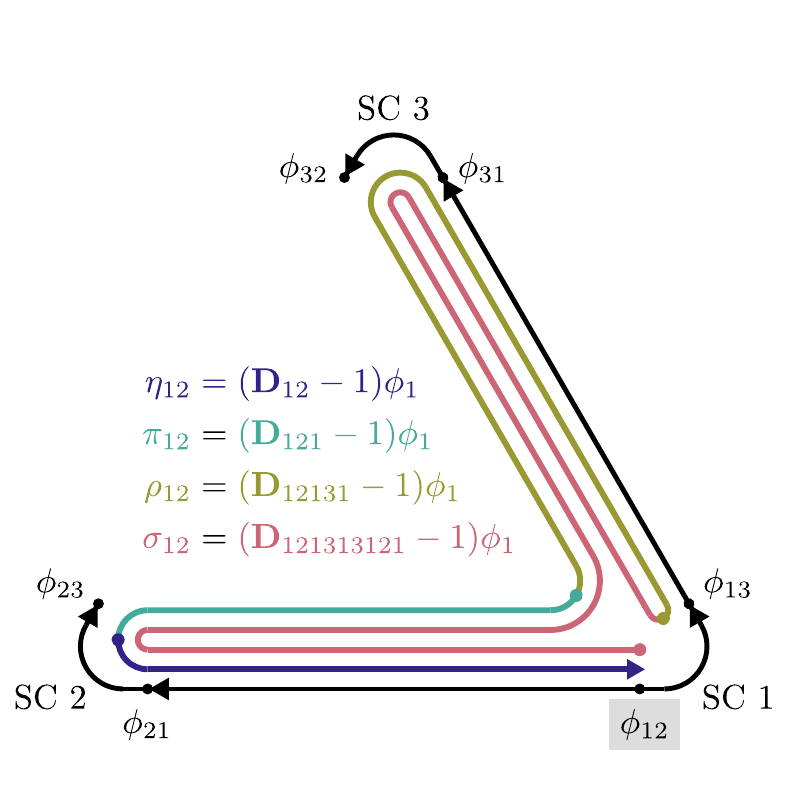}
    \caption{Illustration of intermediary variables defined in \cref{eq:TDIfactorization} and locking configuration N1-12. The intermediary variables are depicted as an arrow representing the synthesized photon path of the long arm. They are incrementally build up from the previously defined ones such that the last variable $\sigma$ is represented by the entire path. Additionally, the chain of locked lasers for the locking configuration N1-12 is shown. The primary laser is highlighted by the grey box.}
    \label{fig:intervars}
\end{figure}

In this section we present how the effects stemming from commutators and post-processing appear in the second-generation \gls{tdi} combination $X_2$. In order to optimize numerical precision and to save computational cost, we calculate $X_2$ in several stages using the following intermediary variables. To start with, the variable $\eta$ is constructed from the inter-spacecraft and reference interferometers. This step reduces the number of lasers from six to three. Then, the variables $\pi$, $\rho$ and $\sigma$ are constructed from $\eta$ by building round trip interferometers of increasing complexity. We thus have
\begin{subequations}
    \begin{align}
        \eta_{ij} &= 
        \begin{cases*}
            \mathrm{isi}_{ij} - \fDelay{ij} \frac{\mathrm{rfi}_{jk} - \mathrm{rfi}_{ji}}{2} & if $\epsilon_{ijk} = 1$, \\
            \mathrm{isi}_{ij} + \frac{\mathrm{rfi}_{ik} - \mathrm{rfi}_{ij}}{2} & if $\epsilon_{ijk} = -1$,
        \end{cases*} \label{eq:eta}\\
        \pi_{ij} &= \eta_{ij} + \fDelay{ij}\eta_{ji}, \\
        \rho_{ij} &= \pi_{ij} + \fDelay{iji}\pi_{ik}, \\
        \sigma_{ij} &= \rho_{ij} + \fDelay{ijiki}\rho_{ik}.
    \end{align}
\label{eq:TDIfactorization}
\end{subequations}
\Cref{fig:intervars} provides an illustration of the intermediary variables as they synthesize two-beam interferometers with a long and a short arm. The long arm is depicted as an arrow propagating around the \gls{lisa} constellation.
Using the variable $\sigma$ we can finally express the second-generation Michelson combination $X_2$ as
\begin{equation}
    X_2 = \sigma_{13} - \sigma_{12}.
    \label{eq:michelson}
\end{equation}
The contracted post-processing delays $\fdelay{i_1 \cdots i_N}$ appearing in \cref{eq:TDIfactorization} are applied in two steps. First, the nested delay $d_{i_1 \cdots i_N}$ is calculated, and then, the fractional delay filter is applied. The nested delay can be calculated using the recursive operation\footnote{The interpolation error arising in this operation might also be relevant as it contributes to the ranging error. To suppress it below the modulation noise level we use an appropriate interpolation orders (order 5 seems sufficient) in our numerical studies.}
\begin{equation}
    d_{i_1 \cdots i_N} = d_{i_1 i_2} + \fdelay{i_1 i_2} d_{i_2 \cdots i_N} .
\end{equation}
Alternatively, contracted post-processing delays can be decomposed into atomic delay operations
\begin{equation}
    \fdelay{i_1 \cdots i_N} \rightarrow \fdelay{i_1 i_2} \cdots \fdelay{i_{N-1} i_N} .
\end{equation}
Note that the expressions on the left and right of the arrow are not equivalent and will produce a different overall interpolation residual in \gls{tdi}, see below. Whether a single contracted delay operator or its decomposition in atomic delays performs better depends on the numerical values of the delays. As a general rule, delaying a time series by a small fractional delay is favorable over delaying a time series with a large one, since the interpolation residual vanishes for an integer delay.

To obtain an explicit expression for the residual in $X_2$ we use the factorization given in \cref{eq:TDIfactorization,eq:michelson} together with the expressions for the inter-spacecraft and reference interferometers in \cref{eq:phasemeter_comm,eq:rfi}. As the couplings discussed in \cref{sec:commutator_residuals,sec:processing_residuals} apply only to either the beatnote frequency offsets or the beatnote phase fluctuations, we split up the interferometers into those contributions. For brevity, we define $a_{ij} = \dot{\mathrm{isi}}_{ij}^o$ and $b_i = \dot{\mathrm{rfi}}_{ij}^o = -\dot{\mathrm{rfi}}_{ji}^o$\footnote{This relation holds for $ij = 12, 23, 31$ as adjacent reference interferometers track the same beatnote up to a sign.} as in~\cite{Hartwig:2022wcs}. The beatnote phase fluctuations are given by
\begin{subequations}
    \begin{align}
        \mathrm{isi}_{ij}^\epsilon &= \delay{ij} \sample\filter p_{ji} - \sample\filter p_{ij} + [\sample\filter, \delay{ij}] p_{ji}, \\
        \mathrm{rfi}_{ij}^\epsilon &= \sample\filter p_{ik} - \sample\filter p_{ij}.
    \end{align}
\end{subequations}
Here, $p_{ij}$ denotes the phase noise of laser $ij$. In the following we write as a short-hand notation $\bar p_{ij} = \sample\filter p_{ij}$.

Next, we replace each occurrence of the post-processing delay $\fDelay{ij}$ by the relation given in \cref{eq:offline_delay_residual}. Here, we neglect all second-order terms in the residuals. Doing so, we find that $X_2$ breaks down into the following residuals
\begin{equation}
    X_2 = \delta X_2^{[\sample\filter,\delay{}]} + \delta X_2^\fdelay{} + \delta X_2^{\Delay{}} + \delta X_2,
\label{eq:TDIresiduals}
\end{equation}
where the origin of the first three terms on the right-hand side is described in \cref{sec:commutator_residuals,sec:processing_residuals}. Additionally, the residuals are propagated through \gls{tdi} and therefore appear multiple times with different delays. The fourth term in \cref{eq:TDIresiduals} is the usual delay commutator describing the arm length mismatch in the virtual interferometer.

In the following we discuss the constituents of \cref{eq:TDIresiduals} individually. To simplify the expressions we assume equal arms and denote the transfer that is common to all residuals as
\begin{align}
    \tdi &\equiv (1 - \delay{}^4) (1 - \delay{}^2), \\
    \tilde\tdi &\equiv 16 \sin^2(2\pi f d) \sin^2(4\pi f d).
\end{align}

The residuals induced by filtering and decimation enter in the inter-spacecraft interferometer (cf., \cref{eq:phasemeter_comm}) and are thus only propagated through \gls{tdi}. This is consistent with~\cite{Bayle:2018hnm}, and they read
\begin{equation}
    \begin{split}
        \delta X_2^{[\sample\filter,\delay{}]} &= \tdi \{[\sample\filter,\delay{13}] p_{31} + \delay{} [\sample\filter,\delay{31}] p_{13} \\
        &\quad\;\; -[\sample\filter,\delay{12}] p_{21} - \delay{} [\sample\filter,\delay{21}] p_{12}\}.
    \end{split} \label{eq:X_sample_filter}
\end{equation}

The interpolation residual depends on the factorization scheme used to compute the \gls{tdi} variables. Using the factorization from \cref{eq:TDIfactorization} with ``contracted delays'', we obtain
\begin{equation}
    \begin{split}
        \delta X_2^{\fdelay{}} &= \tdi \delay{}[\bm{\Delta}_{12} \bar p_{21} - \delay{} \bm{\Delta}_{31} \bar p_{13} - \bm{\Delta}_{13} \bar p_{31} \\
        &\quad + \delay{} (\bm{\Delta}_{13}\!+\!\bm{\Delta}_{31}\!-\!\bm{\Delta}_{12}\!+\!\bm{\Delta}_{121}\!-\!\bm{\Delta}_{131}) \bar p_{12}] .
    \end{split} \label{eq:X_interpolation_residual_fac}
\end{equation}
If one instead performs the interpolation with ``atomic delays'' (i.e., turning $\fDelay{i_1 \cdots i_N}$ into $\fDelay{i_1 i_2}\cdots\fDelay{i_{N-1} i_N}$), one obtains
\begin{equation}
    \begin{split}
        \delta X_2^{\fdelay{}} &= \tdi \delay{}\{\bm{\Delta}_{12} \bar p_{21} + \delay{} \bm{\Delta}_{21} \bar p_{12} \\
        &\qquad\qquad -\delay{} \bm{\Delta}_{31} \bar p_{13} - \bm{\Delta}_{13} \bar p_{31}\} .
        \label{eq:X_interpolation_residual_atomic}
    \end{split}
\end{equation}

To compute the \gls{tdi} residual induced by the ranging error we need a description  of the ranging measurement and any additional processing employed to reduce this ranging error. In \gls{lisa}, two additional modulations on the laser beam are used to measure the inter-satellite range: the \gls{prn} modulation (absolute ranging) and the clock sidebands. The ranging processing described in \cref{app:ranging_processing}, which is mostly adopted from~\cite{Hartwig:2022wcs}, combines both measurements. The resulting ranging estimates $\hat d_{ij}$ inherit a bias $B_{ij}$ from the \gls{prn} measurements and a stochastic term from the sideband measurements. Therefore, the ranging error reads
\begin{equation}
    r_{ij}(t) = B_{ij} - \delay{ij}M_j(t) + M_i(t),
    \label{eq:lisa_ranging_error}
\end{equation}
where $M_i$ denotes the modulation noise on left-handed \glspl{mosa}. 
The resulting residual $\delta X_2^{\Delay{}}$ can be decomposed into the component originating from the ranging bias $\delta X_2^{B}$ and the modulation noise $\delta X_2^{M}$. These two contributions are consistent with~\cite{Hartwig:2022wcs} and read
\begin{subequations}
    \begin{align}
        \begin{split}
            \delta X_2^{B} &= \tdi \{ \delay{}^2 B_{31} \dot{\bar p}_{13} + \delay{} B_{13} \dot{\bar p}_{31} \\
            &\qquad - \delay{}^2 B_{21} \dot{\bar p}_{12} - \delay{} B_{12} \dot{\bar p}_{21}\},
        \end{split}
        \label{eq:X_bias} \\
        \begin{split}
            \delta X_2^{M} &= \tdi \{ ( a_{13} - a_{12} + (1 - \delay{}^2) b_1) M_1 \\
            &\qquad - \delay{} a_{21} M_2 + \delay{} a_{31} M_3 \},
        \end{split}
        \label{eq:X_modulation}
    \end{align}
    \label{eq:X_ranging_error}
\end{subequations}
where $\dot{\bar p}_{ij}$ is the time derivative of the filtered and decimated laser phase fluctuations.

Let us finally give the expression for the usual \gls{tdi} delay commutator that fundamentally limits laser noise reduction. Due to the flexing of the \gls{lisa} constellation the round-trip times of any synthesized two-beam interferometer are not exactly identical. Hence, laser noise in the two beams does not cancel but enters into the \gls{tdi} combination proportional to the arm length mismatch $\Delta d$. For the second-generation Michelson variable $X_2$, the residual reads
\begin{align}
    \delta X_2 &= [\delay{13121}, \delay{12131}] \sample\filter \phi_{12} \\
    &\simeq -\Delta d_{X_2} \delay{131212131} \frac{\mathrm{d}}{\mathrm{d}t}\sample\filter \phi_{12}.
    \label{eq:X_delay_commutator}
\end{align}
In the second line we use the property that a delay commutator acts like a derivative, as already described in~\cite{Shaddock:2003dj,Cornish:2003tz,Bayle:2021mue,Hartwig:2022wcs}. The arm length mismatch in a second-generation Michelson interferometer is given by
\begin{equation}
    \Delta d_{X_2}(t) = d_{131212131}(t) - d_{121313121}(t).
\end{equation}
\Cref{eq:X_delay_commutator} can be further split up into a deterministic out-of-band drift and an in-band component by plugging in \cref{eq:nu}. As the deterministic part, we recover\footnote{As the phase ramp produced by the \si{\tera\Hz} central laser frequency is unaffected by filtering and decimation we can drop both operations.}
\begin{align}
    \delta X_2^o = -\Delta d_{X_2} \delay{131212131} \nu_{12}^o,
    \label{eq:X_delay_commutator_offset}
\end{align}
where the travel time difference can be efficiently computed from the time delays $d_{ij}$ as described in \cref{app:differential_doppler_shift}. Additionally, a model for the absolute laser frequency $\nu_{12}^o$ must be provided to subtract the trend. The stochastic in-band component of the delay commutator is characterized by its \gls{psd}
\begin{equation}
        S_{\delta X_2}(f) = \left(\Delta d_{X_2}\right)^2 (2\pi f)^2 \, \tilde\sample\tilde\filter S_p(f).
\end{equation}
For this derivation we have assumed that $\Delta d_{X_2}$ is constant. In reality, the amplitude of residual laser noise is modulated due to orbital dynamics governing the motion of the three spacecraft. We expand the travel time difference $\Delta d_{X_2}$ up to second order in velocity $\dot d_{iji}$ and up to first order in acceleration $\ddot d_{iji}$ to derive a good approximation of $\Delta d_{X_2}$~\cite{Bayle:2021mue},
\begin{equation}
    \begin{split}
        \Delta d_{X_2} &= (d_{131} \dot d_{121} - d_{121} \dot d_{131})(\dot d_{121} + \dot d_{131}) \\
        &\qquad - (d_{131} \ddot d_{121} - d_{121} \ddot d_{131})(d_{121} + d_{131}),
    \end{split}
\end{equation}
which is of the order \SI{e-12}{\s}~\cite{Hartwig:2022wcs}.

In previous studies~\cite[e.g.,][]{Nam:2022rqg}, it was demonstrated that laser locking does not affect the coupling of path-length noises in \gls{tdi} combinations. At first this seems counter-intuitive as locked lasers generate echoes of any path-length noise imprinted on the reference beam. However, those echos are canceled out in \gls{tdi} since any in-band component in all six laser phases is, by construction, strongly suppressed by the algorithm.

The couplings described above introduce residual laser phase in the \gls{tdi} combinations. Thus, laser noise and path-length noises imprinted on the laser will enter the combination. However, the effect of the latter is subdominant, because laser noise dominates the residuals. The impact of locking is thus only relevant for laser noise. In the following sections we first describe the residuals assuming six independent lasers, i.e., when each laser is locked to an individual cavity; then, we derive the same residuals assuming the standard locking configuration N1-12\footnote{Note that in this work, we shall use the locking configuration naming convention introduced in \cite{Bayle:2022okx}.} introduced in~\cite{heinzel2018}.

\subsection{Six independent lasers}
\label{sec:six_lasers}

For the case of six lasers stabilized to their individual cavities, we assume that their in-band phase noises $p_{ij}$ are independent. Thus, we can compute the total \gls{psd} as the sum of the \glspl{psd} of each laser's contribution.

Using \cref{eq:psd_comm_filter_delay} and the upper bound given in \cref{eq:psd_comm_sample_delay_sim}, it is easy to compute the \gls{psd} of \cref{eq:X_sample_filter} corresponding to the coupling of the filter-delay and decimation-delay commutator. It reads
\begin{subequations}
    \begin{align}
        S_{\delta X_2}^{\sample[\filter,\delay{}]}(f) &= 4\tilde\tdi\tilde\sample\left(\bar{\dot d}^2 \left|\frac{1}{2\pi}\frac{\mathrm{d}\tilde h_\filter(f)}{\mathrm{d}f}\right|^2 S_{\dot p}(f)\right),
        \label{eq:X_psd_comm_filter_delay}
        \\
        S_{\delta X_2}^{[\sample,\delay{}]\filter}(f) &\le 16 \tilde\tdi \cdot \sum_{n=1}^\infty (\tilde\filter S_p)^{(n)}(f),
        \label{eq:X_psd_comm_sample_delay}
    \end{align}
    \label{eq:X_psd_comm_sample_filter_delay}
\end{subequations}
with the effective squared delay derivative
\begin{equation}
    \bar{\dot d}^2 = \frac{\dot d_{12}^2 + \dot d_{21}^2 + \dot d_{13}^2 + \dot d_{31}^2}{4} .
\end{equation}

Next, we investigate the \gls{psd} of the interpolation residual contribution. For six independent lasers we choose to use atomic delays over contracted delays because it results in much simpler couplings. Furthermore, we assume the worst-case interpolation error in order to derive an upper bound on the interpolation error. We use \cref{eq:interpolation_error_psd,eq:X_interpolation_residual_atomic} and find
\begin{equation}
    S_{\delta X_2}^{\fdelay{}}(f) \le 4\tilde\tdi \tilde{\bm{\Delta}} \tilde\sample\tilde\filter S_p(f) .\label{eq:X_psd_interpolation_residual_wc}
\end{equation}
Here, $\tilde{\bm{\Delta}}$ (without indices) represents the worst case interpolation error coupling.

Finally, using \cref{eq:ranging_residual_psd,eq:X_ranging_error}, the contribution from the ranging error yields a \gls{psd} equal to~\cite{Hartwig:2022wcs}
\begin{equation}
    \begin{split}
    S_{\delta X_2}^{\Delay{}}(f) &= \tilde\tdi \Big(4\bar B^2 (2\pi f)^2 \tilde\sample\tilde\filter S_p(f) \\
        &\qquad\qquad\qquad + A_{X_2}^M(f) S_M(f)\Big) ,\label{eq:X_psd_ranging_error}
    \end{split}
\end{equation}
with an effective squared bias $\bar B^2$ and a modulating function $A_{X_2}^M(f)$ defined as
\begin{align}
    \bar B^2 &= \frac{B_{12}^2+B_{21}^2+B_{13}^2+B_{31}^2}{4}, \\
    \begin{split}
        A_{X_2}^M(f) &= (a_{12} - a_{13})^2 + a_{21}^2 + a_{31}^2 \\
        &\qquad- 4 b_1 (a_{12} - a_{13} - b_1) \sin^2(2\pi f d) .
    \end{split}
\end{align}

\subsection{Locked lasers}
\label{sec:locked_lasers}
The baseline design of \gls{lisa} foresees locked lasers to ensure that all beatnote frequencies fall into the sensitive bandwidth of the photoreceivers onboard the spacecraft (\SIrange{5}{25}{\mega\Hz}). To achieve this the primary laser is stabilized using a cavity that serves as a frequency reference. The five remaining lasers are frequency offset locked in succession to the primary following a locking topology. Laser locking of one laser is achieved by adjusting the frequency of this laser source so that the beatnote frequency of the locking interferometer follows a predetermined offset frequency $o_{ij}$. The so-called locking conditions for the inter-spacecraft and reference locking interferometer are given by
\begin{subequations}
    \begin{align}
        \dot{\mathrm{isi}}_{ij} &= \dotdelay{ij} \nu_{ji} - \nu_{ij} = o_{ij} \\
        \dot{\mathrm{rfi}}_{ij} &= \nu_{ik} - \nu_{ij} = o_{ij}
    \end{align}
\end{subequations}
where $\nu_{ji}$ and $\nu_{ik}$, respectively, denote the frequencies of the reference lasers and $\nu_{ij}$ is the frequency of the laser that is controlled. As the offset frequencies $o_{ij}$ only have out-of-band components, any locked laser is simply ``echoing'' the incoming phase noise of the reference laser. Therefore, laser noise becomes correlated among the six lasers.

In this section, we take as an example the N1-12 locking configuration as depicted in \cref{fig:intervars}. Here, N1 specifies the locking topology and 12 the index of the primary laser (see~\cite{heinzel2018,Bayle:2022okx} for an overview of the locking topologies). For this particular locking configuration the in-band phase noise of the six lasers is given by
\begin{equation}
    \begin{split}
        \begin{aligned}[t]
            p_{12} &= p, & p_{23} &= \delay{21}p, & p_{31} &= \delay{31}p, \\
            p_{13} &= p, & p_{32} &= \delay{31}p, & p_{21} &= \delay{21}p.
        \end{aligned} \label{eq:locked_lasers}
    \end{split}
\end{equation}
For better readability we drop the index on $p=p_{12}$ denoting the in-band phase fluctuations of the primary laser. We proceed by inserting the expressions of \cref{eq:locked_lasers} into the general expressions for laser noise related residuals listed in \cref{sec:michelson}. To simplify the expressions we make use of the commutator rule
\begin{equation}
    [\sample\filter, \delay{ij}] \delay{jk} = [\sample\filter, \delay{ijk}] - \delay{ij} [\sample\filter, \delay{jk}].
\end{equation}
We note that the following results are very particular to the choice of locking configuration. Moreover, results for $X_2$, $Y_2$ and $Z_2$ no longer exhibit rotational symmetry in the indices as it is broken by laser locking;  this is easily verified from \cref{eq:locked_lasers}.

As an example, we derive the laser noise residuals for the second-generation Michelson $X_2$ variable. In general, we find that expressions simplify. This is due to the fact that the locking configuration N1-12 inherently generates round-trip measurements required to build $X_2$\footnote{For example, the inter-spacecraft measurements $\mathrm{isi}_{12}\sim\pi_{12}$ and $\mathrm{isi}_{13}\sim\pi_{13}$.}. For $Y_2$ and $Z_2$ we find more complicated expressions involving more terms. However, the \gls{tdi} transfer function remains and additional factors only modulate the residual slightly. Those are caused by cross-spectral densities among correlated laser noise residuals.

For the $[\sample\filter, \delay{}]$ commutator we recover
\begin{equation}
    \delta X_2^{[\sample\filter,\delay{}]} = \tdi ([\sample\filter,\delay{121}] - [\sample\filter,\delay{131}]) p . \label{eq:X_locked_comm_sample_filter_delay}
\end{equation}
We split it up further (as in \cref{eq:comm_sample_filter_delay}) into contributions from the filter-delay commutator and sample-delay commutator, respectively. The \gls{psd} of the flexing-filtering coupling is given by \cref{eq:X_psd_comm_filter_delay} with an effective squared delay derivative equal to
\begin{equation}
    \bar{\dot d}^2 = \frac{(\dot d_{121} - \dot d_{131})^2}{4}.    
\end{equation}
For the sample-delay commutator, the upper bound given in \cref{eq:X_psd_comm_sample_delay} is still valid. Furthermore, we note that both residuals vanish non-trivially if the round-trip delays $d_{121}$ and $d_{131}$ become equal (for the flexing-filtering coupling, their derivatives must also be equal).

To simplify the coupling of the interpolation error to laser noise for locked lasers we use contracted delays (see \cref{eq:X_interpolation_residual_fac}). We obtain
\begin{equation}
    \delta X_2^{\fdelay{}} = \tdi \delay{}^2(\bm{\Delta}_{131} - \bm{\Delta}_{121}) \sample\filter p . \label{eq:X_interpolation_residual_locked}
\end{equation}
Again, the upper bound given in \cref{eq:X_psd_interpolation_residual_wc} still holds and the residual vanishes for equal round-trip times $d_{121}$ and $d_{131}$.

Finally, we discuss the residual caused by ranging errors. The contribution coming from modulation noise in \cref{eq:X_modulation} stays unchanged as the laser noise component of the beatnote is not involved. The contribution of the ranging biases in each arm, see \cref{eq:X_bias}, is given by
\begin{equation}
    \delta X_2^{B} = \tdi \delay{}^2 (B_{12} + B_{21} - B_{13} - B_{31}) \sample\filter p.
\end{equation}
Therefore, the \gls{psd} of ranging error contributions is expressed as \cref{eq:X_psd_ranging_error} with an effective squared bias equal to
\begin{equation}
    \bar B^2 = \frac{(B_{12}+B_{21}-B_{13}-B_{31})^2}{4} .
\end{equation}




\subsection{Comparison with numerical simulations}%

\begin{figure*}
    \centering
    \includegraphics{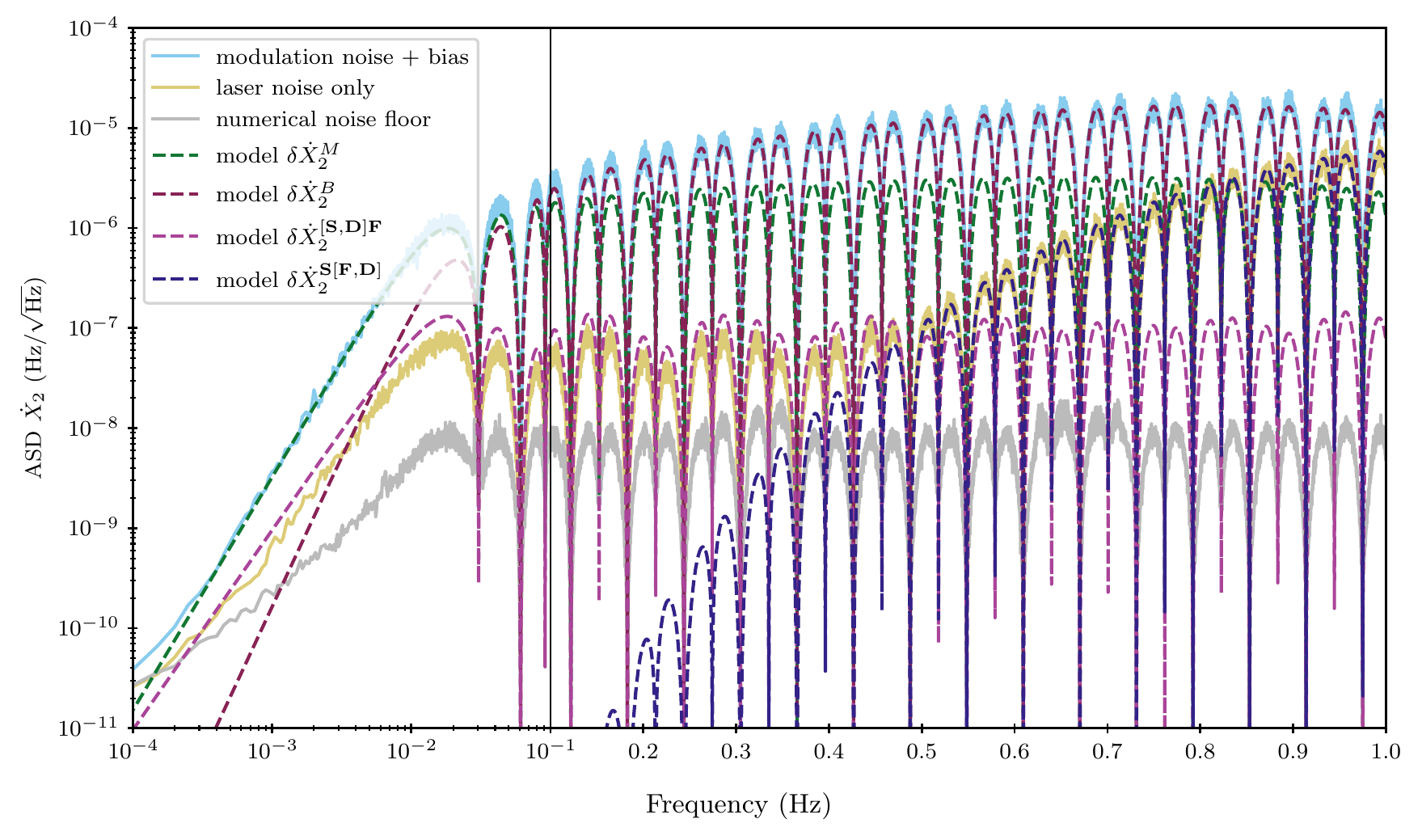}
    \caption{\Glspl{asd} of the second-generation Michelson variable $\dot X_2$ for six independent lasers. The dashed lines correspond to the theoretical predictions (upper bound for aliasing) for the different \gls{tdi} residuals. Solid lines are numerical estimates resulting from the three simulations described in the text. The black vertical at \SI{E-1}{\hertz} divides the x-axis into logarithmically and linearly scaled. Note that the effects of interpolation appears for frequencies larger than \SI{1}{\hertz} and are shown in \cref{fig:michelson_interpolation}.}
    \label{fig:michelson_six_main}
\end{figure*}

\begin{figure*}
    \centering
    \includegraphics{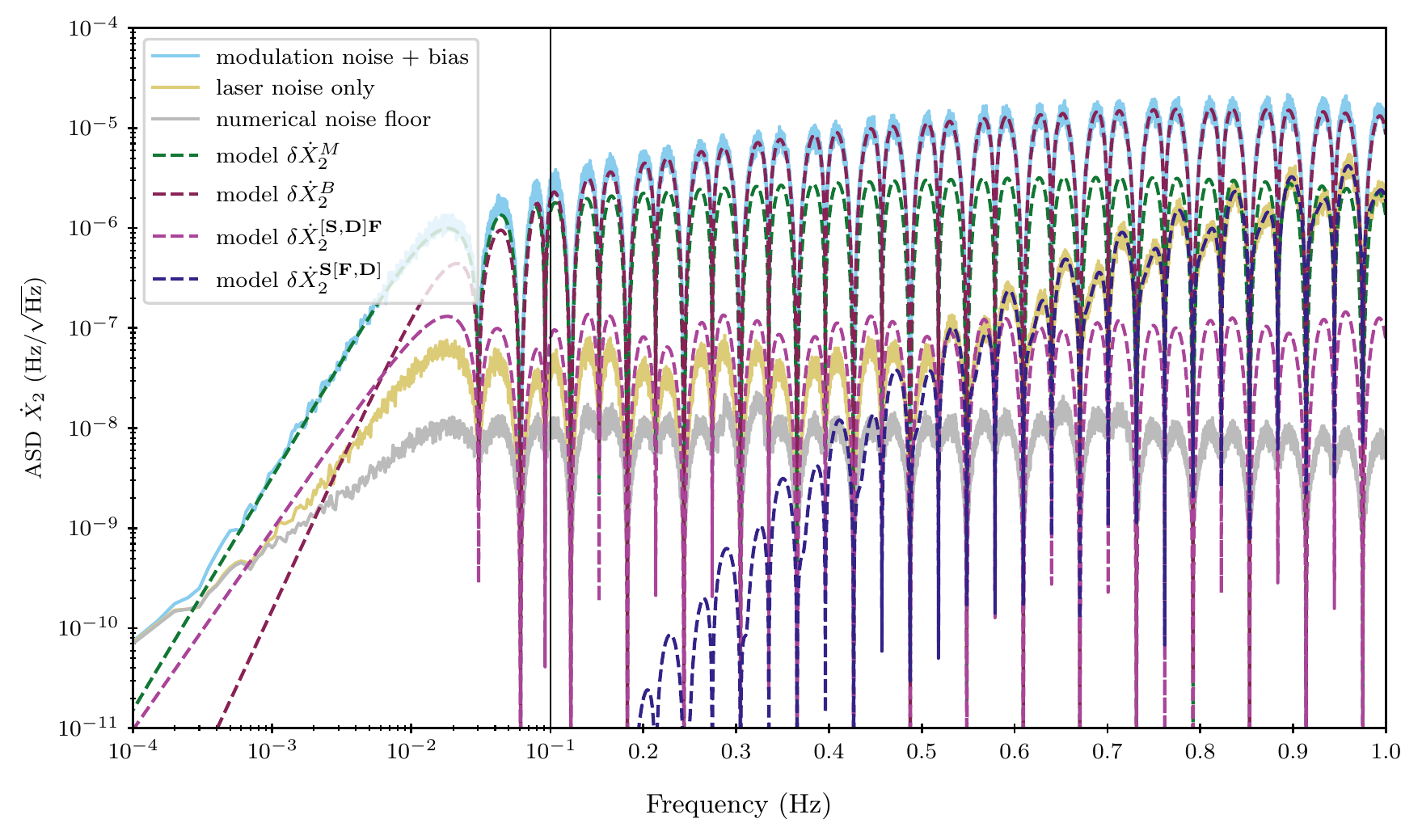}
    \caption{\Glspl{asd} of the second-generation Michelson variable $\dot X_2$ for locked lasers (N1-12). The dashed lines correspond to the theoretical predictions (upper bound for aliasing) for the different \gls{tdi} residuals.  The full lines are numerical estimates resulting from the three simulations described in the text. The black vertical at \SI{E-1}{\hertz} divides the x-axis into logarithmically and linearly scaled. Note that the effects of interpolation appears for frequencies larger than \SI{1}{\hertz} and are shown in \cref{fig:michelson_interpolation}.}
    \label{fig:michelson_N1_main}
\end{figure*}

\begin{figure*}
    \centering
    \includegraphics{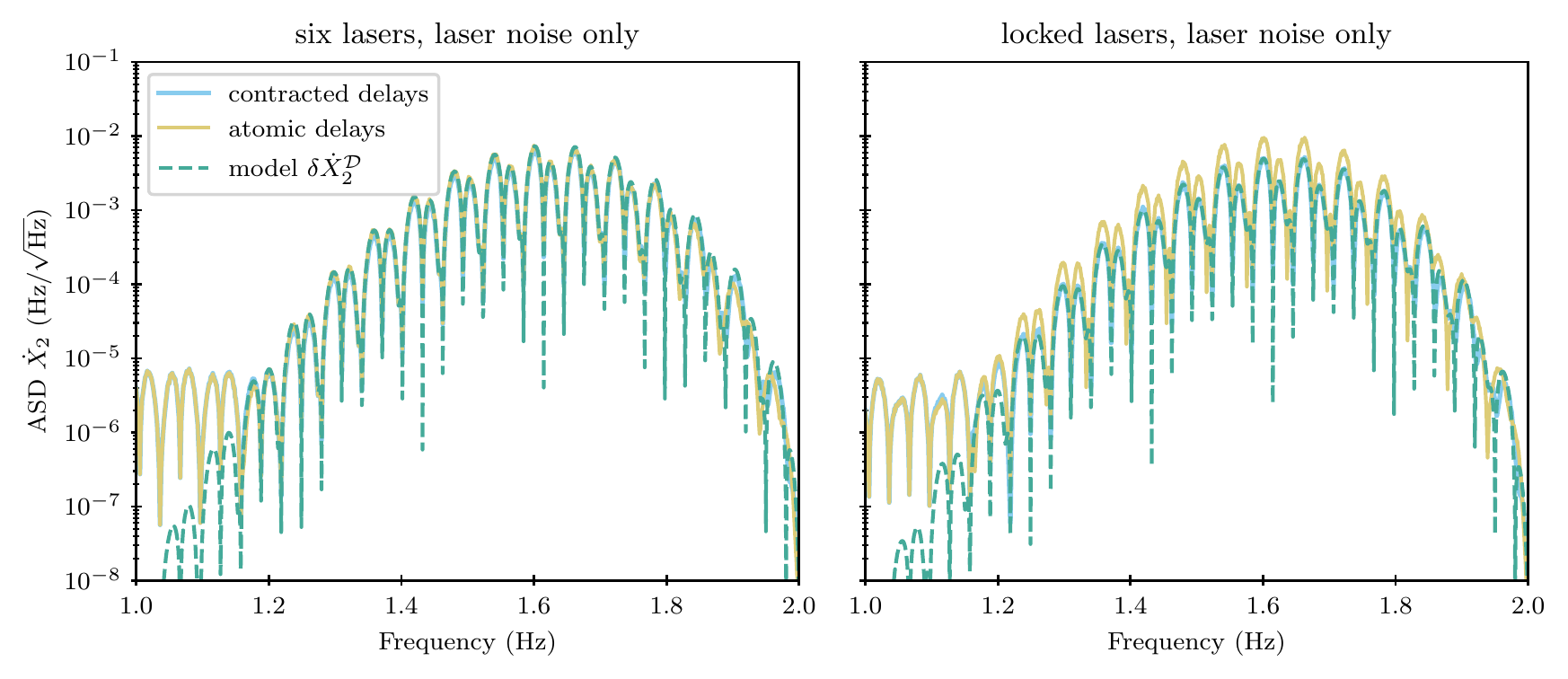}
    \caption{\Glspl{asd} of residual noise in second-generation Michelson variable $\dot X_2$ due to interpolation errors for six independent laser noises (left) and locked lasers (right). The numerical \glspl{asd} using nested (cyan) and flat delay operators (yellow) are compared against (exact) models derived from \cref{eq:X_interpolation_residual_fac} and  \cref{eq:X_interpolation_residual_atomic}, respectively.}
    \label{fig:michelson_interpolation}
\end{figure*}

We now compare the theoretical model described above to simulated \gls{lisa} data, obtained using \texttt{LISA Instrument}~\cite{bayle_jean_baptiste_2022_7071251,Bayle:2022okx} to generate the \gls{lisa} measurements and \texttt{PyTDI}~\cite{staab_martin_2022_6867012} to calculate the second-generation Michelson variables. To compare the couplings described above to the simulated data, individually, we run three simulations with increasing complexity.
In the first simulation, all sources of noise are disabled.  Any residual in the \gls{tdi} variables can be attributed to numerical effects that are due to rounding errors in the simulated floating-point variables. In the second simulation, we add white laser frequency noise with an \gls{asd} equal to \SI{30}{\Hz\per\sqrt{\Hz}}. This gives rise to the residuals caused by the flexing-filtering effect, the coupling of the sampling-delay commutator and interpolation errors. Finally, we introduce modulation noise and ranging biases of the order of \SI{30}{\nano\s} (a factor 10 higher than in \cite{Hartwig:2022wcs} in order to accentuate the coupling) to produce the couplings related to ranging errors. The \gls{asd} of the modulation noise on left-handed \glspl{mosa} is given by
\begin{equation}
    \sqrt{S_M(f)} = \SI{8.3e-15}{\s\per\sqrt{Hz}} \left(\frac{f}{\si{\Hz}}\right)^{-2/3} , \label{eq:modulation_noise}
\end{equation}
while on the right-handed \glspl{mosa} it a factor of 10 larger~\cite{Barke:2015srr}. As already presented in~\cite{Hartwig:2022wcs,Hartwig:2020tdu}, we can obtain an estimate of the pseudo-range $\hat d_{ij}$ that is free of the modulation errors of the right-handed \glspl{mosa}.

All simulations span approximately 3 days in duration with the \gls{lisa} constellation following realistic heliocentric orbits provided by \gls{esa}. Physics is simulated at \SI{16}{\Hz} and the filter presented in \cref{app:filter_design} is used for anti-aliasing before decimating to \SI{4}{\Hz}. Here, the filter design is not optimized using a trade-off between the number of taps and the level of laser noise residual but rather chosen such that the processing residuals are above numerical noises and can be validated against analytical models.


\texttt{LISA Instrument} produces measurements in units of frequency in order to circumvent numerical issues specific to phase units. As a result, the data produced contains the total beatnote frequencies of the inter-spacecraft and the reference interferometers, which have offsets between \SIrange{5}{25}{\mega\Hz}. In addition to the carrier-to-carrier beatnotes, sideband beatnotes that precisely track the delay derivatives are also available.

Prior to any processing all measurements are converted from a 64-bit to an 80-bit floating-point variable. This makes sure that no additional numerical noise is introduced downstream. In the first processing step we extract the high-precision ranging information from the sidebands (see \cref{app:ranging_processing}). Then, the second-generation Michelson variable $X_2$ is calculated using the factorization expression in \cref{eq:TDIfactorization}. Here, we use Lagrange interpolation with $n=62$ coefficients. To convert our expressions to frequency units, all delay operators $\fDelay{ij}$ are replaced by $\dotfDelay{ij}$, which are defined analogously to \cref{eq:dotdelay}. In the last processing step, we subtract any out-of-band drifts from the Michelson combinations to reduce the effect of spectral leakage at DC. This is achieved by computing the differential Doppler shift, as explained in \cref{app:differential_doppler_shift}, and inserting it into the time derivative of \cref{eq:X_delay_commutator_offset}.

The resulting \glspl{asd} of the simulations for six independent and locked lasers are presented in \cref{fig:michelson_six_main,fig:michelson_N1_main,fig:michelson_interpolation}. To match the analytical models at high frequencies (i.e. flexing-filtering effect and interpolation error) with the numerical \glspl{asd} for locked lasers we had to relax the equal-arms assumption and use the expressions for six unequal but constant arms. \Cref{fig:michelson_six_main,fig:michelson_N1_main} show that for the laser-noise-only simulations, the residuals are well explained by the commutator residuals. For both six independent and locked lasers, we plot the same upper bound for aliased laser noise from \cref{eq:X_psd_comm_sample_delay}. We note that the locked case exhibits a slightly reduced noise level, which can be explained by partial cancellation of the two contributions appearing in \cref{eq:X_locked_comm_sample_filter_delay}. Adding modulation noise and ranging bias to the simulation completely dominates the aforementioned effects. Indeed, the residual in $X_2$ is now dominated by the ranging and interpolation residuals, which are again well explained by their analytical counterparts. At very low frequencies (\SIrange{e-4}{e-3}{\Hz}) we observe deviations from the models. Those can explained by the numerical noise floor arising from rounding-errors in the simulation. We also observe this numerical limit in the noiseless simulation shown in grey.

In \cref{fig:michelson_interpolation}, we show the \glspl{psd} of interpolation errors for six independent and locked lasers using either contracted or atomic delays, respectively. As a side note, this residual only affects frequencies outside the \gls{lisa} band ($>\SI{1}{\Hz}$) and thus should be irrelevant from the point of view of the performance of TDI. This being said, the focus of this paper is to correctly model the residuals, which is why we include this computation. In the left hand plot, we show the exact models (assuming six unequal but constant arms) which can be seen to match the numerical results. In the right hand plot, we show a similar plot but for locked lasers.  On the one hand, we find that, in this particular case, it is indeed advantageous to use contracted delays to calculate $X_2$ for locked lasers in the N1-12 locking configuration as it produces a residual that is smaller by approximately a factor of two compared to using atomic delays. On the other hand, for six independent lasers we virtually see no difference. As discussed earlier in this section, there is no general rule for the optimal factorization of delays as it is dependent on their particular value. However, in the worst case scenario ($\epsilon = 0.5$), atomic delays outperform contracted delays for six lasers and vice versa for locked lasers (see \cref{eq:X_interpolation_residual_atomic,eq:X_interpolation_residual_locked}).

\section{Conclusion}%

In this paper, we have presented a comprehensive study of residual laser noise in the \gls{tdi} variables for \gls{lisa}. We have identified two categories of couplings. First, onboard processing steps, namely filtering and decimation, give rise to additive noise in the inter-spacecraft interferometers due to non-commutation with the delay operation. Secondly, the post-processing delays employed to calculate \gls{tdi} combinations only partially mitigate laser noise. This is because this offline computation relies on an interpolation method which induces interpolation errors, and because the offline delays used for \gls{tdi} include ranging errors. For both categories of laser noise couplings, we provide analytical models for the residuals in the second-generation \gls{tdi} combination $X_2$; we validate those models using numerical simulations.

In the existing literature, the flexing-filtering effect~\cite{Bayle:2018hnm} and the coupling of ranging errors~\cite{Hartwig:2022wcs} for six independent lasers are already described and preliminary models of laser noise residuals due to aliasing and interpolation errors are presented in~\cite{Hartwig:2021dlc}. In this study, we remedy the shortcomings of the latter and present all laser noise couplings in a consistent framework. Furthermore, we investigate the impact of laser locking by discussing the example of the locking configuration N1-12. Finally, because we perform \gls{tdi} in total frequency units, we explain the deterministic trend that is present in second-generation \gls{tdi} combinations: differential Doppler shifts in the round-trip paths of the synthesized beams produce a beatnote of a few \si{\milli\Hz}.  This trend depends solely on the out-of-band delays due to orbital dynamics and the \si{\tera\Hz} frequency of the involved laser. It can therefore be computed and removed by appropriately modeling the orbits and the \si{\tera\Hz} frequency evolution. This detrending step is a reversible (the trend can always be added in again) part of pre-processing. It occurs before parameter estimation and reduces spectral leakage in \gls{psd} estimates, a feature which is relevant for the present study.


Contrary to unsuppressed noises~\cite{Nam:2022rqg} (e.g. path length noises), the coupling of laser noise residuals is dependent on the underlying locking configuration. This can be explained by the fact that locked lasers follow the primary laser with configuration-specific time lags, which introduce correlations among all lasers. To be consistent with the existing literature, we first derive analytical models for six independent laser. Then, we repeat the calculation for locked lasers, more specifically the configuration N1-12.
In this configuration, the spacecraft 2 and 3 act as transponders directly sending light back to spacecraft 1. This means the inter-spacecraft interferometer beatnotes recorded on spacecraft 1 already represent the signal combinations $\eta_{12} + \delay{12}\eta_{21}$ and $\eta_{13} + \delay{13}\eta_{31}$ for laser noise, simplifying the expression for the Michelson $X_2$ variable. However, this simplification does not apply to $Y_2$ and $Z_2$. In general, analytic models become more complicated for locked lasers due to the introduced correlations.
Furthermore, we find that generally the worst case scenario in terms of laser noise residuals can be larger for locked lasers than for six independent laser. Hence, future worst case studies should account for all possible locking configurations.


The level of additional noise due to onboard processing (filtering and decimation) is strongly dependent on the design of the anti-aliasing filter. In this study, we used a \gls{fir} filter with a transition band ranging from \SIrange{0.1}{2}{\Hz} so as to relax requirements on filter implementation (the number of taps is reduced to 103 compared to 145 in the standard \texttt{LISA Instrument} implementation) and accentuate the flexing-filtering coupling in the \gls{lisa} band. In theory, the flexing-filtering coupling can be mitigated by flattening the response of the filter in the pass-band on ground. The appropriate design of compensation filters is the subject of on-going efforts in the \gls{lisa} community. Indeed, any aliased noise due to insufficient attenuation in the stop-band cannot be reduced in post-processing and has to be taken care of before decimation.


Laser noise residuals stemming from post-processing delays depend on the interpolation method and the ranging performance. In this paper, we extensively rely on Lagrange interpolation, which has a maximal flat response at DC. Alternative interpolation kernels are given by the family of ``windowed sinc'' kernels~\cite{Shaddock:2004ua} or numerically optimized kernels. These are currently under study. Interpolation kernels of shorter length have smaller computational cost and result in less truncation at the boundaries (where the interpolation kernel does not completely overlap with the data). This problem becomes more critical in the presence of gaps. In this paper, we also studied the impact of contracting processing delays, i.e., combining nested delays first to form a single delay operation. We find that the best delay contraction strategy depends on the locking configuration and the particular numerical values of the delays.


The results presented in this paper should be invariant under the time reference frames the measurements are defined in. Therefore, the general findings should still be valid for measurements sampled according to realistic clocks that are processed using ``Time-delay interferometry without clock synchronization'' presented in~\cite{Hartwig:2022wcs}. Additionally to flexing arms due to orbital dynamics, clock drifts of the order of \SI{e-7}{} become relevant for the flexing-filtering effect. Furthermore, extra care must be taken when extracting the delay estimates from the sideband measurements since measured pseudo-ranges have an in-band component. Here, appropriate compensation filter are crucial since sideband beatnotes are also subject to anti-alias filtering. Any departure from unity in the filter's transfer function will produce additional ranging noise in the delay estimates.


\section*{Acknowledgments}
The authors would like to thank G. Heinzel for the valuable discussions and the feedback on the manuscript.
M.S. and O.H. acknowledge the support of the German Space Agency, DLR. The work is supported by the Federal Ministry for Economic Affairs and Climate Action based on a decision by the German Bundestag (FKZ 50OQ1801 and FKZ 50OQ2301). This work is also supported by the Max-Planck-Society within the LEGACY (``Low-Frequency Gravitational Wave Astronomy in Space'') collaboration (M.IF.A.QOP18098). 
J.-B. B. gratefully acknowledges support from the UK Space Agency via STFC [ST/W002825/1]. M.L.'s and O.H.'s work was supported by the Programme National GRAM of CNRS/INSU with INP and IN2P3 co-funded by CNES. O.H. and M.L. gratefully acknowledge support from the Centre National d'\'Etudes Spatiales (CNES). 

\appendix
\section{Anti-aliasing filter design}
\label{app:filter_design}

\begin{figure}
    \centering
    \includegraphics{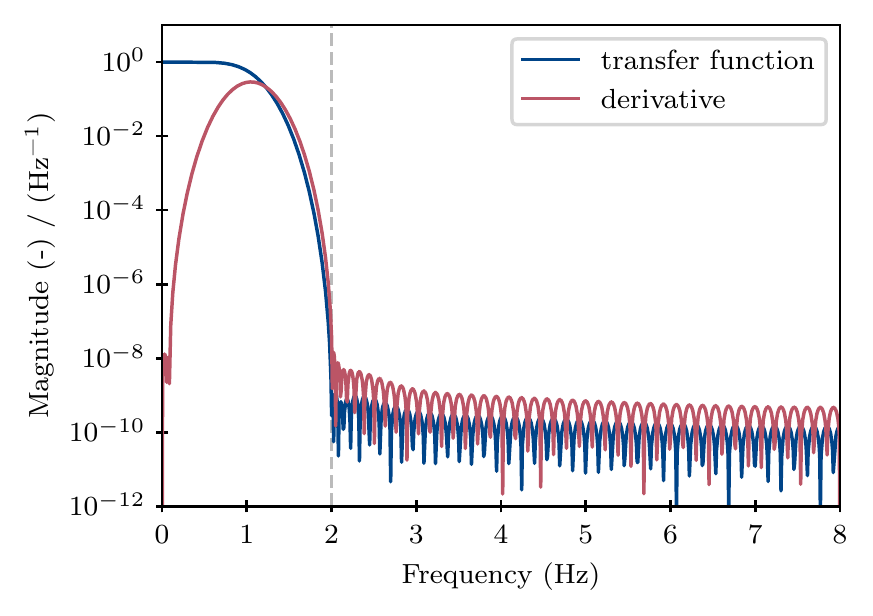}
    \caption{Magnitude of filter transfer function (blue) and its frequency derivative (red). The grey dashed line indicates the cutoff frequency at \SI{2}{\Hz}.}\label{fig:filter_design}
\end{figure}

The main objective of the anti-aliasing filter is to prevent folding of power during decimation. In the numerical simulations performed in this work, we simulate the physics at \SI{16}{\Hz} and decimate to \SI{4}{\Hz}. Therefore, we need to design the anti-aliasing filter such that it cuts off at a Nyquist frequency of \SI{2}{\Hz}.  We use the routines from the \texttt{SciPy} Python package to design a \gls{fir} filter with the so called ``window method''. Here, we assume a pass-band frequency of \SI{0.1}{\Hz}, a stop-band frequency of \SI{2}{\Hz} and a minimum attenuation of \SI{181}{\decibel}. The resulting transfer function and its derivative which is of relevance for the flexing-filtering coupling (see \cref{ssec:flexing_filtering_coupling}) are plotted in \cref{fig:filter_design}.

\section{Ranging processing}
\label{app:ranging_processing}

The ranging processing discussed in this section is mostly adopted from~\cite{Hartwig:2022wcs} and we introduce only minor changes to the algorithm. The main differences are that we write it down in units of phase and reformulate the suppression of modulation noise contributions originating from right-handed optical benches. Additionally, we assume that the sideband phases are read out using a feed-forward scheme. This measure avoids tracking the carrier phase redundantly and also accounts for the difference in modulation frequencies on each \gls{mosa}. Therefore, the sideband phases of the inter-spacecraft and reference interferometer read
\begin{subequations}
    \begin{align}
        \mathrm{isi}_{ij}^\mathrm{sb} &= \delay{ij}\phi_{ji}^\mathrm{m} - \phi_{ij}^\mathrm{m} - (\nu_{ji}^\mathrm{m} - \nu_{ij}^\mathrm{m}) t, \\
        \mathrm{rfi}_{ij}^\mathrm{sb} &= \phi_{ik}^\mathrm{m} - \phi_{ij}^\mathrm{m} - (\nu_{ik}^\mathrm{m} - \nu_{ij}^\mathrm{m}) t,
    \end{align}
    \label{eq:sidebands}
\end{subequations}
where the phase of the modulation is given by
\begin{equation}
    \phi_{ij}^\mathrm{m}(t) = \nu_{ij}^\mathrm{m}\cdot(t + M_{ij}(t)).
\end{equation}
Here, $M_{ij}(t)$ accounts for timing jitter due to modulation noise. Inserting this definition into \cref{eq:sidebands} yields
\begin{subequations}
    \begin{align}
        \mathrm{isi}_{ij}^\mathrm{sb} &= -\nu_{ji}^\mathrm{m} d_{ij} + \delay{ij} \nu_{ji}^\mathrm{m} M_{ji} - \nu_{ij}^\mathrm{m} M_{ij}, \\
        \mathrm{rfi}_{ij}^\mathrm{sb} &= \nu_{ik}^\mathrm{m} M_{ik} - \nu_{ij}^\mathrm{m} M_{ij},
    \end{align}
\end{subequations}
which has a similar algebraic structure to the carrier phases (c.f. \cref{eq:phasemeter}) where the product $\nu_{ij}^\mathrm{m} M_{ij}$ takes the place of the laser phase $\phi_{ij}$. We now use the definition of the intermediary variable $\eta_{ij}$ for the sidebands\footnote{The post-processing delays required to calculate $\eta_{ij}^\mathrm{sb}$ have much more relaxed error requirements than is the case for laser noise cancellation. We can therefore use the delay estimates from the \gls{prn}.} (see \cref{eq:eta}) to cancel modulation noise contributions stemming from right-handed \glspl{mosa},
\begin{equation}
    \eta_{ij}^\mathrm{sb} = -\nu_{ji}^\mathrm{m} d_{ij} + \delay{ij} \nu_{jk}^\mathrm{m} M_j - \nu_{ij}^\mathrm{m} M_i.
\end{equation}
Here, we use the shorthand notation $M_1 = M_{12}$, $M_2 = M_{23}$ and $M_{3} = M_{31}$.

Finally, the variables $\eta_{ij}^\mathrm{sb}$ have to be scaled by the respective modulation frequency and multiplied by $-1$ in order to yield a low noise estimate of the delay
\begin{equation}
    \hat d_{ij} = -\frac{\eta_{ij}^\mathrm{sb}}{\nu_{ji}^\mathrm{m}} = d_{ij} - \delay{ij} \frac{\nu_{jk}^\mathrm{m}}{\nu_{ji}^\mathrm{m}} M_j + \frac{\nu_{ij}^\mathrm{m}}{\nu_{ji}^\mathrm{m}} M_i \label{eq:ranging_estimate} .
\end{equation}
Assuming that modulation frequencies only differ by a fraction of a percent we can approximate the above expression by setting the ratios of frequencies to 1. Doing so, we recover the stochastic component in \cref{eq:lisa_ranging_error}.

In practice, we process data in frequency units. Hence, the above procedure is rewritten in frequency by taking a global time derivative. By doing so, \cref{eq:ranging_estimate} yields an accurate measurement of the delay derivative $\dot{\hat d}_{ij}$, which then needs to be integrated to recover the delay itself, required for \gls{tdi}. The integration constant is derived from the \gls{prn} ranging measurement~\cite{Hartwig:2022wcs}.

\section{Differential Doppler shift}
\label{app:differential_doppler_shift}

As explained in \cref{sec:michelson}, any \gls{tdi} combination representing a virtual two-beam interferometer does not cancel the laser phase perfectly for flexing arms but is limited by a residual given by the delay commutator. The origin of this residual is the travel time difference $\Delta d$ between the two virtual beams. The deterministic component of this residual can be calculated and subtracted from the \gls{tdi} observable (see \cref{eq:X_delay_commutator_offset}). Here, we present an efficient scheme to calculate $\Delta d$. First we recognize that the delay $d_{ij}$ can be written as
\begin{equation}
    d_{ij}(t) = t - \delay{ij}t,
\end{equation}
which has the same algebraic structure as $\eta_{ij}$ up to a sign. Here, the time argument $t$ takes the place of the laser phase $\phi_i$ and $\phi_j$ (c.f. \cref{eq:phasemeter,eq:TDIfactorization}). Using $\eta_{ij} = d_{ij}$ as inputs to \gls{tdi} yields the travel time difference $\Delta d$ as
\begin{equation}
    X = -[\delay{A}, \delay{B}] t = d_{AB} - d_{BA} = \Delta d,
\end{equation}
where $AB$ and $BA$ denote the paths of the counter-propagating beams of an arbitrary \gls{tdi} combination $X$ representing a two-beam interferometer.

As processing is performed in frequency units in this paper we are more interested in the derivative of $\Delta d$. To avoid numerical problems we thus operate on the delay derivatives $\dot d_{ij}$ directly and form $\Delta\dot d = \dot X$ following the procedure explained in~\cite{Bayle:2021mue}.

\section{Coupling of ranging noise to laser noise}
\label{app:ranging_laser_noise_coupling}
In \cref{sec:processing_residuals}, we neglect the coupling of the stochastic component of the ranging error to laser noise as it appears to be much weaker compared with the coupling to the \si{\mega\Hz} beatnote frequency. However, for the sake of completeness and as it becomes relevant in processing pipelines where one removes the phase ramp operates on the fluctuations directly we present the coupling mechanism below.

To suppress all other laser noise couplings in the final \gls{tdi} combination we consider a setup where we have already removed the phase ramp from the interferometric measurements such that they only track the differential phase noise $p_{ij}$ of the six lasers,
\begin{subequations}
    \begin{align}
        \mathrm{isi}_{ij}(t) &= \delay{ij} p_{ji}(t) - p_{ij}(t), \\
        \mathrm{rfi}_{ij}(t) &= p_{ik}(t) - p_{ij}(t).
    \end{align}
    \label{eq:phasemeter_laser_noise}
\end{subequations}
For simplicity we omit the anti-aliasing filtering and decimation that was considered in \cref{eq:phasemeter}.

Next, we insert \cref{eq:phasemeter_laser_noise} into \cref{eq:eta} where the post-processing delay $\Delay{}$ is used and only accounts for a stochastic ranging error $r(t)$. We can express $\eta_{ij}$ for the left and right-handed \glspl{mosa} as
\begin{equation}
    \eta_{ij} = \Delay{ij} p_j - p_i - (\Delay{ij} - \delay{ij}) p_{ji} . \label{eq:eta_laser_noise}
\end{equation}
Here, we use the short-hand notation $p_1 = p_{12}$, $p_2 = p_{23}$ and $p_3 = p_{31}$. We recognize that the last term in \cref{eq:eta_laser_noise} is already a laser noise residual and we neglect higher order couplings in the following. Then, we use the intermediary variables, $\pi_{ij}$, $\rho_{ij}$ and $\sigma_{ij}$ defined in \cref{eq:TDIfactorization,eq:michelson} to find the total laser noise residual in the second-generation Michelson combination $X_2$. It consists of the residual in \cref{eq:eta_laser_noise} that is propagated through \gls{tdi} as well as the commutator of post-processing delay operators
\begin{equation}
    \begin{split}
        [\Delay{13121}, \Delay{12131}] &= [\delay{13121}, \delay{12131}] \\
        &\quad+ (\Delay{131212131} - \delay{131212131}) \\
        &\quad- (\Delay{121313121} - \delay{121313121})
    \end{split}
\end{equation}
applied to $p_1$. Here, we split the commutator into the ``usual'' delay commutator and two additional terms that produce further laser noise residuals. For equal arms, the full residual in $X_2$ reads
\begin{equation}
    \begin{split}
        \delta X_2^\Delay{} &= \tdi \{\delay{} (r_{31} \delay{} \dot p_{13}) + (r_{13} \delay{} \dot p_{31}) \\
        &\qquad - \delay{} (r_{21} \delay{} \dot p_{12}) - (r_{12} \delay{} \dot p_{21})\} \\
        &\quad + \tdi \{r_{12} + \delay{} r_{21} - r_{13} - \delay{}r_{31}\} \cdot \delay{}^8 \dot p_1,
    \end{split} \label{eq:X_ranging_noise}
\end{equation}
where we have used \cref{eq:ranging_residual} and have approximated nested ranging noise as
\begin{subequations}
    \begin{align}
        r_{ijikikiji} &\simeq r_{iji} + \delay{}^2 r_{iki} + \delay{}^4 r_{iki} + \delay{}^6 r_{iji} , \\
        r_{iji} &\simeq r_{ij} + \delay{} r_{ji}.
    \end{align}    
\end{subequations}
Finally, we compute the \gls{psd} of \cref{eq:X_ranging_noise} by assuming that all laser and ranging noise terms are uncorrelated and have identical noise properties. We find 
\begin{equation}
    \begin{split}
        S_{\delta X_2}^{\Delay{}}(f) &= 4 \tilde\tdi (S_r(f') * S_{\dot p}(f'))(f) \\
        &\qquad+ 4 (\tilde\tdi S_r(f') * S_{\dot p}(f'))(f),
    \end{split}
\end{equation}
where we have neglected any cross terms between the first two lines and the last line in \cref{eq:X_ranging_noise}. The $*$ sign denotes convolution (in frequency domain) which stems from the time domain products of ranging and laser noise contributions.

\bibliography{references}

\end{document}